\documentclass{nature}
\pdfoutput=1
\usepackage{graphicx}

\usepackage{aas_macros}
\usepackage{amsmath}
\usepackage{amssymb}
\usepackage{gensymb}
\usepackage{makecell}
\usepackage{longtable}
\usepackage[font={sf}]{caption}
\usepackage{hyperref}
\usepackage[displaymath,mathlines]{lineno} 
\newcommand{\Sun}{\odot}

\title{Testing the universality of free fall by tracking a pulsar in a stellar triple system}

\author{Anne M. Archibald$^{1,2}$, Nina V. Gusinskaia$^1$, Jason W. T. Hessels$^{1,2}$, Adam T. Deller$^3$, David L. Kaplan$^4$, Duncan R. Lorimer$^{5,6}$, Ryan S. Lynch$^{7,6}$, Scott M. Ransom$^8$ \& Ingrid H. Stairs$^9$}

\begin{document}

\maketitle

\begin{affiliations}
 \item Anton Pannekoek Institute for Astronomy, University of Amsterdam, Science Park 904, 1098 XH Amsterdam, The Netherlands
 \item ASTRON, Netherlands Institute for Radio Astronomy, Postbus 2, 7990 AA, Dwingeloo, The Netherlands
 \item Centre for Astrophysics and Supercomputing, Swinburne University of Technology, P.O. Box 218, Hawthorn, VIC 3122, Australia
 \item University of Wisconsin-Milwaukee, P.O. Box 413, Milwaukee, WI 53201, USA
 \item Department of Physics and Astronomy, West Virginia University, P.O. Box 6315, Morgantown, WV 26506, USA
 \item Center for Gravitational Waves and Cosmology, Chestnut Ridge Research Building, Morgantown, WV 26505, USA
  \item Green Bank Observatory, P.O. Box 2, Green Bank, WV 24944, USA
 \item National Radio Astronomy Observatory, Charlottesville, VA 22903, USA
 \item Department of Physics and Astronomy, University of British Columbia, 6224 Agricultural Road, Vancouver, BC V6T 1Z1, Canada
\end{affiliations}

\begin{abstract}
Einstein's theory of gravity, general relativity\cite{eins16}, has passed stringent tests in laboratories, elsewhere in the Solar Sytem, and in pulsar binaries\cite{will14}. Nevertheless it is known to be incompatible with quantum mechanics and must differ from the true behaviour of matter in strong fields and at small spatial scales. A key aspect of general relativity to test is the strong equivalence principle (SEP), which states that all freely falling objects, regardless of how strong their gravity, experience the same acceleration in the same gravitational field. Essentially all alternatives to general relativity violate this principle at some level\cite{deru11}. For example, string theories generically predict a scalar field, the dilaton, that affects the motion of falling bodies\cite{dp94}. Previous direct tests of the SEP are limited by the weak gravity of the bodies in the Earth-Moon-Sun system\cite{hm18} or by the weak gravitational pull of the Galaxy on pulsar-white dwarf binaries\cite{zdw+18}. PSR~J0337+1715 is a hierarchical stellar triple system\cite{rsa+14}, where the inner binary consists of a millisecond radio pulsar in a $1.6$-day orbit with a white dwarf. This inner binary is in a $327$-day orbit with another white dwarf. In this system, the pulsar and the inner companion fall toward the outer companion with an acceleration about $10^8$ times greater than that produced by falling in the Galactic potential, and the pulsar's gravitational binding energy is roughly $10\%$ of its mass. Here we report that in spite of the pulsar's strong gravity, the accelerations experienced by it and the inner white dwarf differ by a fraction of no more than $2.6\times 10^{-6}$ ($95\%$ confidence level). We can roughly compare this to other SEP tests by using the strong-field Nordtvedt parameter $\hat\eta_N$. Our limit on $\hat\eta_N$ is a factor of ten smaller than that obtained from (weak-field) Solar-System SEP tests\cite{hm18,gmg+18} and a factor of almost a thousand smaller than that obtained from other strong-field SEP tests\cite{zdw+18}. 
\end{abstract}

%Although this hierarchical triple is stable in the long term\cite{rafi14}, three-body interactions produce complicated orbital motions\cite{rafi14}, measurement of which allows the individual stellar masses and nearly all orbital parameters to be determined\cite{rsa+14}. The pulsar's rotation is extremely stable, and orbital complications from tidal effects and stellar winds are negligible (Methods).

We observed PSR~J0337+1715 with the Westerbork Synthesis Radio Telescope (WSRT), the Robert C. Byrd Green Bank Telescope (GBT), and the William E. Gordon telescope at the Arecibo Observatory (AO). We have over 800 observations spanning approximately six years, which total about 1200 hours on source. During each observation we folded (summed in time) the rotationally modulated radio signal from the pulsar according to a preliminary model for the pulsar's sky position, spin rate, and orbital motion. We recorded flux density as a function of rotational phase, radio frequency, and time. We processed these observations using techniques developed for precision pulsar timing\cite{nab+15} (Methods). In this process, each folded profile was compared to a standard template (Figure~\ref{fig:template}) to determine how early or late the pulses arrived compared to our reference model. We averaged the data in time and frequency; most observations are averaged into roughly 20-minute integrations with 20-MHz bandwidth. This resulted in roughly 27,000 multi-frequency pulse time-of-arrival measurements (TOAs), with a formal weighted root-mean-square uncertainty of $1.0\,\mu\text{s}$ (for individual telescope data sets: Arecibo $0.4\,\mu\text{s}$, GBT $1.3\,\mu\text{s}$, WSRT $1.6\,\mu\text{s}$).

To accommodate the complex three-body interactions in this system, we modelled the orbits by directly integrating the equations of motion\cite{rsa+14}. To allow testing general relativity, we chose equations of motion that include parametrized post-Newtonian (PPN)\cite{wn72} interactions between bodies. This framework allows essentially all gravitational theories to be approximated to first post-Newtonian order. If we forbid preferred-frame and preferred-location effects as well as non-conservation of momentum, theories in this framework are parametrized by $\beta$ and $\gamma$. Both $\beta$, which measures the non-linearity of gravity, and $\gamma$, which measures the degree to which space-time is curved by gravity, take the value $1$ in general relativity. We chose a point-particle Lagrangian that permits arbitrarily strong gravity internal to the bodies and parametrized post-Newtonian interactions between them\cite{nord85}. We then used computer algebra\cite{msp+17} to construct equations of motion. Each orbit was specified by an initial system configuration at Modified Julian Date (MJD) 55920.0 (2011 Dec 25 00:00:00 UTC). The evolution of this configuration was governed by $\beta$, $\gamma$, and the SEP violation parameter $\Delta$. Following Damour and Sch\"afer\cite{ds91} we define $\Delta = m_G/m_I - 1$, the fractional difference between the pulsar's inertial ($m_I$) and gravitational ($m_G$) masses.  The SEP is satisfied only if $\Delta = 0$.

\begin{figure}[htbp]
\centerline{\includegraphics[width=183mm]{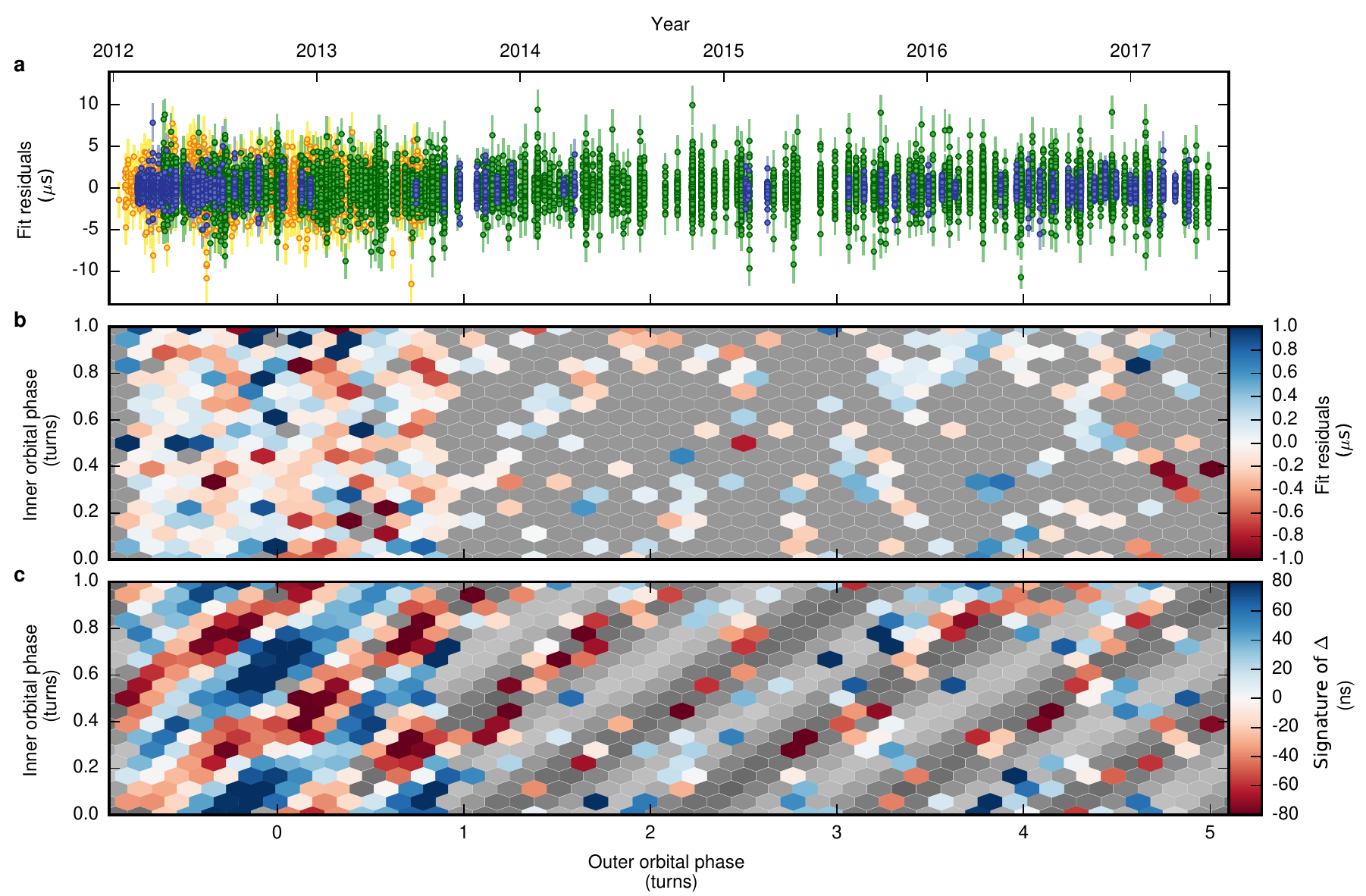}}
\caption{\label{fig:residuals}\textbf{Difference between pulsar time of arrival measurements and model (fit residuals).} \textbf{a}, Fit residuals as a function of time measured in years (top axis) as well as outer orbital phase (bottom axis). Error bars are $1 \sigma$.  Orange points are from WSRT data; green points are from GBT; blue points are from Arecibo. The observational cadence was higher in earlier years.  \textbf{b},  Fit residuals from (a) binned by inner orbital phase and outer orbital phase.  \textbf{c}, Signature of $\Delta$, that is: the pattern of fit residuals that result from introducing $\Delta =  2.6\times 10^{-6}$ but fitting a model with $\Delta$ fixed to zero. Blue and red hexagons show our actual observational sampling, whereas the grayscale hexagons simulate a completely uniform sampling.}
\end{figure}

% Fitting the orbit
Our fitting procedure simulated orbits for trial sets of parameters. Once an orbit had been simulated, we used a linear least-squares fitting process to measure parameters such as pulsar spin period and offset from a reference sky position (Methods). We repeated this for many orbits to search the space of parameters for the best fit, residuals from which are shown in Figure~\ref{fig:residuals}. We also computed numerical derivatives of the orbit with respect to each parameter. This process gave us best-fit values and formal uncertainties on all parameters. We did not meaningfully constrain the parametrized post-Newtonian parameters $\beta$ and $\gamma$, for which the posterior distributions are indistinguishable from the prior distributions (which were based on Solar System observations\cite{bit03,gmg+18}). In contrast, $\Delta$ was substantially constrained by our observations and analysis procedure.

\begin{figure}[htbp]
\centerline{\includegraphics[width=89mm]{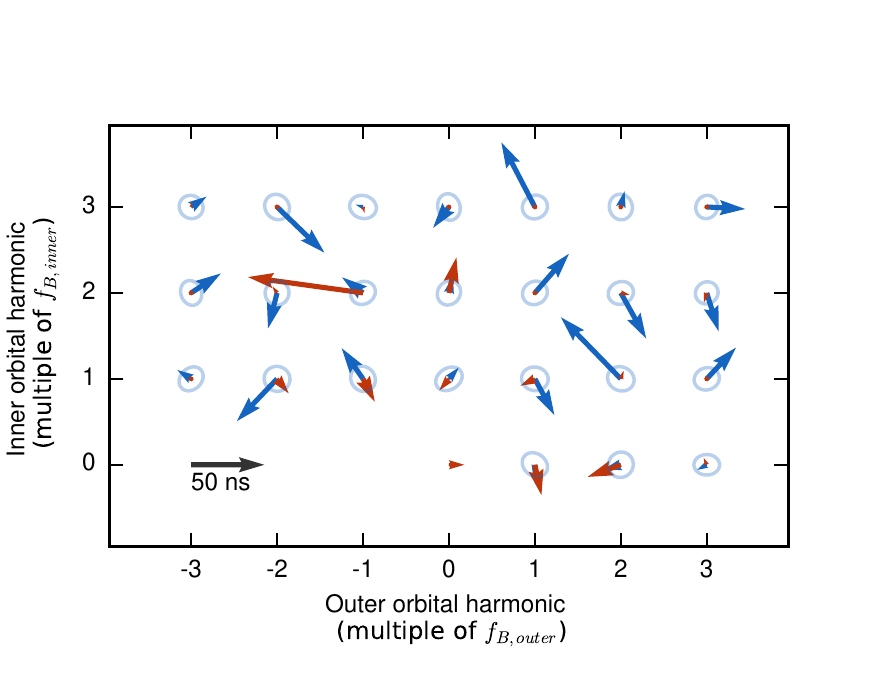}}
\caption{\label{fig:arrowplot}\textbf{Quasi-Fourier representation of fit residuals.} Each arrow length and direction represents, respectively, the amplitude and phase of a sinusoid. For reference, the length of the black arrow corresponds to an amplitude of $50\,\text{ns}$ in the pulse times-of-arrival for a given frequency. Ellipses represent $1\sigma$ arrow lengths coming from the formal uncertainties on our data. Blue arrows represent residuals from the fit. Red arrows represent the signature of an SEP violation, $\Delta=2.6 \times 10^{-6}$, as in Figure~\ref{fig:residuals} c. The longest red arrow is $78\,\text{ns}$.}
\end{figure}

% Results of fitting
The results of our fitting process appear in Tables~\ref{tab:solutionfixed}, \ref{tab:solutionfitted}, and \ref{tab:solutioninferred}. We measure $\Delta = -1.1\times 10^{-6}$ with a formal 1-$\sigma$ uncertainty of $2\times 10^{-7}$. We caution that these formal uncertainties do not include systematic effects such as excess delays caused by the variable solar wind (our line of sight to PSR~J0337+1715 passes within $2.1\degree$ of the Sun every year) or refractive variations in the scattering time (on the order of $30\,\text{ns}$); such effects can be strongly correlated between measurements and can thus significantly affect best-fit values in spite of being much smaller than the formal uncertainties on the pulse arrival times.

% Systematic uncertainties
To obtain a realistic limit on $\Delta$, we carried out a systematics analysis procedure on the residuals from our fit. The key idea was to look at the ``signature'' of a non-zero $\Delta$, that is, the effect on an orbit of introducing a non-zero $\Delta$ and then fitting for all other parameters (as in Figure~\ref{fig:residuals}~c). This signature can be understood from a theoretical point of view: the differential acceleration introduced by a non-zero $\Delta$ shifts the inner binary orbit toward the outer companion.  In the residuals from an operation that fits all parameters except $\Delta$, this produces a sinusoid with frequency $2f_{\text{inner}}-f_{\text{outer}}$ (where $f_{\text{inner}}$ and $f_{\text{outer}}$ are the inner and outer orbital frequencies, respectively; Methods). We can compute this signature of $\Delta$ numerically using our orbit simulator; see Figure~\ref{fig:residuals}~c. It is a sinusoidal variation of pulse arrival times whose amplitude is $30\,\text{ns}$ when $\Delta$ is $10^{-6}$. The fact that the signature has a simple sinusoidal form suggested that, if we wanted to understand how systematics might impact the measured value of $\Delta$, we should look at sinusoids with similar frequencies $kf_{\text{inner}}+lf_{\text{outer}}$, where $k$ and $l$ are integers. Figure~\ref{fig:arrowplot} shows the results of fitting many such sinusoids to the residuals from our best fit. The distribution of coefficients of these sinusoids implies a $1\sigma$ scatter of $22\,\text{ns}$ on the component corresponding to the signature of $\Delta$ (Methods); in combination with the best-fit value of $\Delta$, we obtain $\Delta = (-1.09 \pm 0.74)\times 10^{-6}$, which is consistent with zero at the $2\sigma$ level. This corresponds to a $95\%$ upper limit of $|\Delta|<2.6\times 10^{-6}$.

\begin{table}[htbp]
\begin{center}
{\small
\begin{tabular}{lccccc}
Description of fit & \makecell[c]{$\Delta \pm$ unc.\\ $\times 10^{-6}$}  &  \makecell[c]{stat. unc. \\$\times 10^{-6}$} & \makecell[c]{ ampl. sign.$\Delta$\\ (ns)} & \makecell[c]{syst.unc\\ (ns)} \\
\hline
\multicolumn{5}{c}{Primary fit:}\\
\hline
\makecell[l]{Observatories: AO, GBT, WSRT \\ Frequency band: $\nu_c \sim 1400\,\text{MHz}$ \\ DM fit interval: one year \\ EoM: 1st order PN, $\Delta \neq 0$ } & $-1.1  \pm  0.7$ & $0.2$ & $33$ & $22$\\
\hline                                                                               
\multicolumn{5}{c}{Alternate physical models:}\\                                     
\hline                                                                               
EoM: Newtonian, $\Delta \neq 0$  & $ 0.7  \pm  4.1$ & $0.2$ & $22$ & $132$\\
EoM: 1st order PN of GR, i.e. $\Delta \equiv 0$ & $-$ & $-$ & $-$  &  $23$\\
\hline
\multicolumn{5}{c}{Subsets of data:}\\
\hline
\multicolumn{5}{l}{Single observatory fits (using $\nu_c \sim1400\, \text{MHz}$ data):}\\
\makecell[l]{AO (5355 TOAs, 0.6$\mu s$ WRMS)} & $-0.98 \pm  2.5$ &  $0.3$  &  $25$ & $64$\\
\makecell[l]{GBT (18487 TOAs, 1.5$\mu s$ WRMS)} & $ 0.04 \pm  1.2$  &  $0.4$  &  $1$  & $35$\\                                                                          
\makecell[l]{WSRT (3268 TOAs, 1.8$\mu s$ WRMS)}& $-2.1 \pm  2.5$ &  $1.3$  &  $47$ & $55$\\ 
\multicolumn{5}{l}{Frequency band additions:}\\
\makecell[l]{+AO $430\,\text{MHz}$ + WSRT $350\,\text{MHz}$} & $-2.1\pm  1.5$ & $0.2$ & $67$ &  $47$   \\                                                           \hline                                                                               \end{tabular}                                                                       }                                                                                 
\end{center}
\caption{\label{tab:runs}\textbf{Values of $\Delta$, formal uncertainties, and estimates of systematic errors, from various fit approaches.} The $\Delta$ constraint quoted in the main text is the result of the Primary fit approach. Each other fit in the table differs in exactly one respect, either using an alternate physical model or a different subset of our data.  All uncertainties are 1-$\sigma$. The third column gives the amplitude in nanoseconds of the signature of $\Delta$ found in our pulse arrival times, while the fourth column gives the 1-$\sigma$ amplitude of the systematics. Here EoM -- equation of motion, 1st PN -- the first post-Newtonian approximation, GR -- general relativity, $\nu_{c}$ -- central observing frequency, WRMS -- weighted root mean square of the residuals. 
}
\end{table}

% Tweaking the fitting procedure
To test the robustness of our method, we explored the effect of fitting various subsets of our data, and the influence of handling physical effects in different ways. The values we report in Table~\ref{tab:solutionfitted} are calculated using the best method we found; the results from other approaches are presented in Table~\ref{tab:runs}. In each alternative approach, the limits we obtain on $\Delta$ are compatible but less constraining.

Our result is a direct test of the SEP: in the gravitational pull of the outer white dwarf, the pulsar and the inner white dwarf experience accelerations that differ fractionally by $|\Delta| < 2.6 \times 10^{-6}$ ($95\%$). For comparison, the most similar previous test is based on the pulsar-white dwarf binary PSR~J1713+0747 falling in the gravitational pull of the Galaxy; this constrains $|\Delta|<2\times 10^{-3}$ ($95\%$) in a physically similar situation\cite{zdw+18} but its sensitivity is limited by the low acceleration due to the Galaxy's gravity, which is a factor $10^{-8}$ that in our system. We are therefore able to improve on the previous limit by almost three orders of magnitude. An SEP violation in either of these tests would arise from gravitational phenomena in the interior of a neutron star, one of the strongest-curvature environments accessible to observation\cite{bps15}. 
% Check Ingrid's limit is 2 sigma

\begin{figure}[htbp]
\centerline{\includegraphics[width=\textwidth]{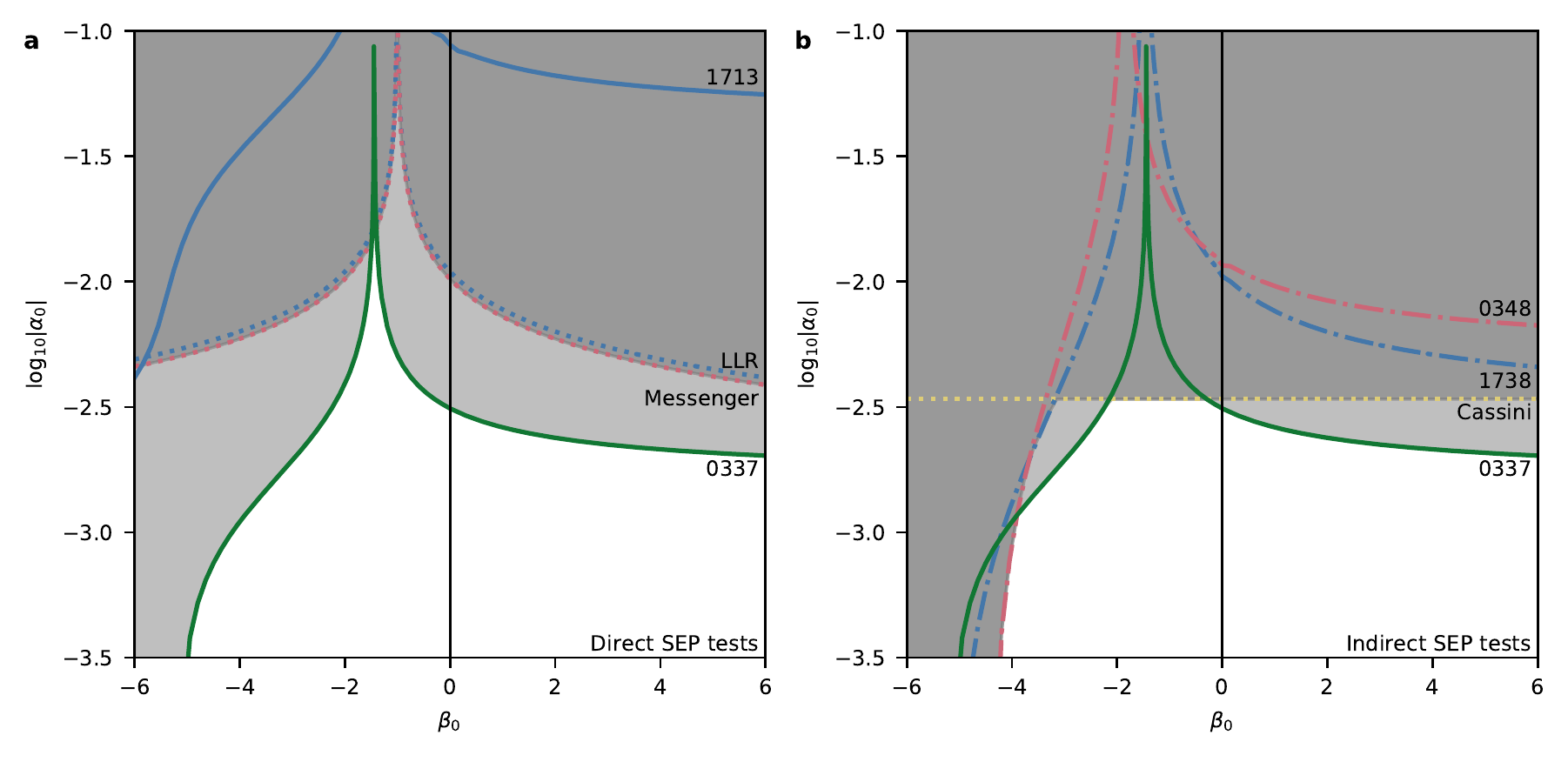}}
\caption{\label{fig:alpha-beta}\textbf{Constraints on quasi-Brans-Dicke theories of gravity.} These theories are parametrized by $\alpha_0$ and $\beta_0$. Existing constraints rule out theories in the dark grey region. The constraint derived in this work, labelled ``0337'', additionally rules out theories in the light grey region. This diagram assumes a very stiff equation of state for neutron stars and is therefore somewhat conservative about neutron-star tests. \textbf{a}, Comparison with existing direct SEP tests (``1713'': the wide binary PSR~J1713+0747\cite{sfl+05}, ``LLR'': lunar laser ranging\cite{hm18} and ``Messenger'': Mercury-orbit\cite{gmg+18} SEP tests). \textbf{b}, Comparison with indirect SEP tests (``Cassini'': Cassini Shapiro delay\cite{bit03} and ``1738'' and ``0348'': dipole gravitational wave limits from neutron-star/white-dwarf binaries\cite{fwe+12,afw+13}) using the assumed theories of gravity.}
\end{figure}

To understand the theoretical implications of our result --- or to compare our result to other, perhaps indirect, tests of the SEP --- we need to select a framework parametrizing alternative theories of gravity. The most common such framework is the parametrized post-Newtonian formalism\cite{wn72}; this parametrizes most alternative theories to first post-Newtonian order. While the interactions between bodies in the PSR~J0337+1715 system are adequately described at this order, the interior of the pulsar is a strong-field region, and it is precisely this region that may cause the pulsar's gravitational mass to differ from its inertial mass. The parametrized post-Newtonian framework is thus not sufficient to describe our result. 
% De-acronym

Although other strong-field frameworks are available\cite{bbc+15} (Methods), we choose the family of quasi-Brans-Dicke theories. These theories, inspired in part by Mach's principle\cite{bd61}, add a scalar field, $\phi$, to general relativity. Within the family, $\alpha_0$ and $\beta_0$ are parameters that select a particular theory (note that this $\beta_0$ is different from the parametrized post-Newtonian parameter $\beta$). These theories permit, for example, the gravitational constant $G$ measured in a Cavendish experiment to depend on the local value of $\phi$. While Solar System experiments are able to constrain $\alpha_0$, for large negative values of $\beta_0$ the phenomenon of ``spontaneous scalarization'' occurs in neutron stars\cite{de92}, allowing the value of $\phi$ inside them to be of order unity regardless of the weak-field behaviour.  Thus quasi-Brans-Dicke theories are well constrained by pulsar experiments, and several key results are summarized, along with our constraint, in Figure~\ref{fig:alpha-beta}. In particular, note that this allows us to compare our own strong-field SEP test with the weak-field lunar laser ranging test\cite{hm18}, the weak-field light-bending test based on the Cassini mission\cite{bit03}, as well as pulsar tests placing upper limits on the emission of gravitational dipole radiation\cite{fwe+12,afw+13}. For most $\beta_0 \gtrsim -4$ our result provides the strongest upper limit on $\alpha_0$ and hence the most stringent constraint on how gravity can deviate from the predictions of Einstein's general relativity. We also dramatically improve upon all other direct tests of the SEP.

\clearpage
\begin{methods}
% Make figure and table numbering reflect being in the online Methods
\renewcommand\thefigure{E.\arabic{figure}}
\setcounter{figure}{0}
\renewcommand\thetable{E.\arabic{table}}
\setcounter{table}{0}

\subsection*{Precision timing}

Arecibo observations were taken with the L-band wide receiver, which has a dual linear polarization feed, and an $800$-MHz band was recorded with the Puerto-Rican Ultimate Pulsar Processing Instrument (PUPPI). GBT observations were taken with the L-band receiver, which has dual linear feeds, and an $800$-MHz band was recorded with the Green-Bank Ultimate Pulsar Processing Instrument\cite{drd+08} (GUPPI). WSRT observations were taken with the Multi-Frequency Front End receivers, which have dual linear feeds, a single tied-array beam on the sky was formed (using phase and polarization calibration determined by the observatory), and a $160$-MHz band was recorded with the Pulsar Machine II\cite{kss08} (PuMa II). All observations were coherently dedispersed\cite{hr75}.

Although we generally follow best practices developed by the pulsar timing array community\cite{nab+15}, PSR~J0337+1715 has a few unusual features that force us to adopt additional, special techniques. Unknown additional features, or known features we were not able to compensate for completely, introduce systematic structure in our residuals. Our systematics analysis procedure serves to estimate their impact on the key parameter $\Delta$, and our reported uncertainty includes the estimated impact of this systematic structure. 
% Cite Verbiest et al. 2016?

The full model describing the pulsar's motion is too complicated to use in real-time observing. We therefore observed while folding 10-second integrations using the pulsar period predicted from a two-non-interacting-Keplerian model\cite{rsa+14} (\texttt{BTX}) or a single-Keplerian-orbit model with varying parameters (\texttt{BTX} also) that is understood by the standard pulsar timing tool TEMPO. These simplified models predict pulse phases that can differ from the observed phases by a substantial fraction of a pulse period towards the end of our observing span. It is therefore necessary to correct the folded archives by phase-shifting the recorded profiles to match the predictions of the full model. This ensures that when we averaged archives into 20-minute spans, they were already aligned so that no further time smearing occurred. At the same time, we were able to compute the phase drift of the observing model within each 10-second integration. It is impossible to correct for this smearing; its amplitude is typically around $200\,\text{ns}$ and can be as large as $1500\,\text{ns}$. More problematically, the model errors can easily be correlated with inner or outer orbital phase, possibly in the same way as the signature of the SEP violation we are looking for. This may explain some of the systematics we detect, and we recommend that future observations be carried out with the more accurate short-term folding models we currently use to realign archives. We plan to release a bundle of such short-term ephemerides covering at least the next few years.

\begin{figure}[htbp]
\centerline{\includegraphics[width=89mm]{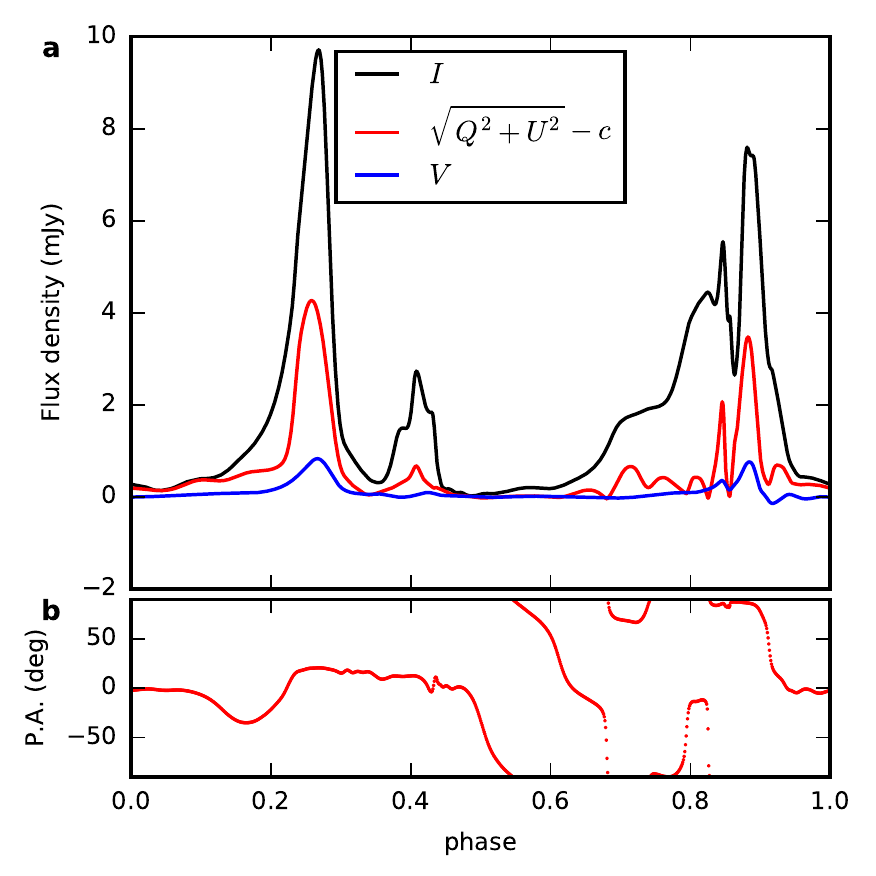}}
\caption{\label{fig:template}\textbf{Template pulse profile used for timing.} This is based on the average $1300$--$1900$~MHz pulse profile from the Green Bank Telescope observation on modified Julian day 56412. The Stokes IQUV data have been smoothed by a wavelet-based algorithm implemented in \texttt{psrsmooth} from the software package PSRCHIVE. \textbf{a}, total intensity (I), linear (Q,U), and circular (V) polarization after correcting for Faraday rotation. \textbf{b}, polarization angle at the centre frequency of the observation. Note that the linear polarization (in red) at some phases is responsible for almost half the flux density and that this pulse profile has complicated polarization structure. Offsets have been added to $I$ and to $\sqrt{Q^2+U^2}$ to ensure that $I^2\geq Q^2+U^2+V^2$.}
\end{figure}
% Cite psrchive and/or software list

The pulse profile from PSR~J0337+1715 includes substantial linear polarization, varying as a function of pulse phase (Figure~\ref{fig:template}). Since all of our telescopes directly measure orthogonal pairs of polarizations,  reconstructing the total intensity profile depends on accurate polarimetric calibration. The WSRT undergoes polarization calibration as part of the tied-array beam-forming process. Calibration data, including a feed and dish model, is available for the GBT, in addition to the noise diode scans we took before each observation. Unfortunately, no feed and dish model is available for Arecibo, and we found that in spite of our use of diode scans, in some observations we were not able to calibrate the Arecibo polarimetry. Specifically, we have examples of Arecibo observations where even with the best available calibration, the reconstructed Stokes I profile differs substantially in shape from the standard template observation; in these cases a suitable polarization transformation is able to match the template to the observation. We therefore adopted a technique similar to ``matrix template matching''\cite{stra06}: when we compare each observation to our accurately calibrated polarimetric template (shown in Figure~\ref{fig:template}), we fit for an arbitrary Mueller matrix, an offset in each of the Stokes IQUV parameters, and a phase shift, transforming the template IQUV to match the observed IQUV values. This process renders our pulse arrival times largely insensitive to polarization calibration and also allows the pulsar's polarization structure to constrain the timing, yielding a 15\% improvement in fit uncertainties compared to a fit using only the total intensity. 

Standard practice in precision pulsar timing is to take low-frequency (for example 430~MHz) observations quasi-simultaneously with each high-frequency (typically 1400~MHz) observation to better constrain dispersion measure variations. Our Arecibo observations were taken in this mode, and on some days WSRT data was acquired at 350~MHz. Unfortunately, we found that if we included this low-frequency data (using a template based on a bright Arecibo 430~MHz observation), our estimate of the systematics in the post-fit timing residuals became measurably worse. This may be the result of interstellar scintillation and scattering: at 1400~MHz, we observe scintillation with a typical frequency structure of 5~MHz. This predicts\cite{lr99,akhs14} a scattering tail of 30~ns, varying by a factor of roughly two on months time scales due to refractive scintillation. This is comparable to the size of the signals we are looking for. While such scintillation is a minor systematic effect on the 1400~MHz data, this time scale is predicted to increase as the negative fourth power of observing frequency\cite{akhs14}, giving a scattering time scale on the order of 8~$\mu$s at 430~MHz, certainly large enough to complicate our use of low-frequency data. We therefore omitted use of these low-frequency observations in our primary fit (but see Table~\ref{tab:runs} for an evaluation of their impact on our result, if included).

Our observations primarily record frequencies $1100$--$1900\,\text{MHz}$. We expect the intrinsic profile of the pulsar to vary as a function of frequency across this range. Nevertheless, we use the single pulse profile template shown in Figure~\ref{fig:template} for all observations in this band. We therefore expect there to be a modest frequency-dependent but time-independent time shift in our data. To compensate for this, we fixed the dispersion measure and fit for a delay that is a polynomial function of the logarithm of frequency\cite{nab+15}. Using the $F$ test we found that four terms were sufficient to model this variability. We therefore include four parameters in our timing model to describe this frequency variation.

Because the ecliptic latitude of PSR~J0337+1715 is only $2.1\degree$, every March our line of sight to the pulsar passes very close to the Sun. The solar wind then contributes potentially significant extra dispersion measure to these observations. While we did fit for an idealized solar wind model each year (see below), we know that the solar wind is time-variable and not spherically symmetric. We therefore excised all data for which the line of sight passed within $5\degree$ of the Sun; this keeps the predicted excess delays due to the solar wind below a few microseconds. Our solar-wind fitting should remove the majority of this, and our systematics estimation should account for the residual effects on our estimate of $\Delta$.

Finally, in this very large collection of 818 observations, a few will inevitably have been corrupted by observer error, telescope malfunctions, or radio-frequency interference. We therefore constructed a summary plot for each observation showing pulse profile versus time and frequency, smearing within an observing sub-integration, and timing residuals relative to the short-term ephemeris used to align the observation. We examined these by eye so we could excise part or all of any problematic observation. In addition to the standard automatic interference excision provided by the program \texttt{paz} from PSRCHIVE, we found it necessary to manually excise interference from 65 observations and to completely discard 17.

\subsection*{Timing model}

\begin{figure}[htbp]
\centerline{\includegraphics[width=89mm]{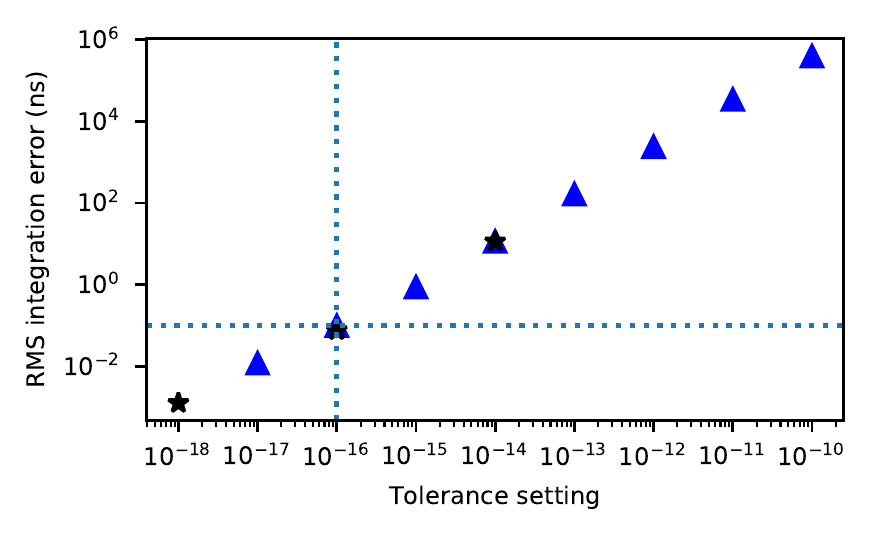}}
\caption{\label{fig:tolerance}\textbf{Timing model truncation error.} This plot shows the RMS arrival-time error caused by the orbital integrator's finite time steps, as a function of the tolerance parameter. The vertical dotted line corresponds to the value used for all other solutions in this work; the RMS error from this source is below 0.1 ns. Blue triangles are calculations done in hardware 80-bit floating point; black stars are calculations done in software 128-bit floating point, which are much slower to compute. To estimate the errors in this plot, we computed a fiducial solution with 128-bit precision and tolerance parameter $10^{-22}$ and compared all other solutions to this.}
\end{figure}

Traditional pulsar timing models rely on formulae expressing Keplerian, or parametrized post-Keplerian, orbits. No such formulae are known that can handle the classical three-body interactions that occur in this system. We therefore implemented our timing model by directly integrating the equations of motion with a Bulirsch-Stoer integrator\cite{bs66}, using root-finding methods on the integrator's dense output to compute the pulsar proper time at which each received pulse was emitted. This orbital modelling has only finite accuracy, limited both by the step size of the differential equation integrator and by the numerical precision with which the millions of steps are accumulated. We addressed truncation error by using an adaptive step-size integrator with a tolerance parameter; we adjusted this tolerance parameter to obtain a negligible truncation error of approximately $0.1$ ns (Figure~\ref{fig:tolerance}). Round-off error we addressed by using 80-bit floating-point to carry out our integrations; we are able to switch to 128-bit floating point as a cross-check, but our machines do not have hardware support for 128-bit floating point, so these high-precision integrations use software floating-point routines and are roughly a factor of 50 slower. Fortunately, by comparing with 128-bit test runs, we found that 80-bit calculations provided small enough round-off error, and we therefore used these for all calculations.
% Dense output reference?

The parameters describing a hypothetical timing solution fall into two categories. Some parameters, for example outer binary period, affect the pulsar's orbit, requiring a new orbit to be simulated when they are changed. Other parameters, for example pulsar spin frequency, can be determined by a linear least-squares fit once the non-linear parameters have been set. Thus optimization can be carried out in two nested stages, one a non-linear downhill optimizer and the other a simple linear least-squares solver. For Bayesian computations, we are able to operate on the non-linear parameters alone by analytically marginalizing over the linear parameters. This analytical marginalization simply amounts to using the linear least-squares best-fit values for the linear parameters and adding a correction to the log-probability computed from the linear least-squares fit matrices. The linear parameters are as follows: For parametrizing the pulsar's spin, we fit for pulsar spin frequency and frequency derivative. For astrometric parameters, we began with the published values of position and distance\cite{rsa+14}, we set proper motion to zero, and we fit for offsets from these values as linear parameters: position offsets, parallax error, and proper motion. We also fit for instrumental delays between telescopes, and we fit for a time-independent but frequency-dependent delay due to profile variations with frequency (see above). Finally, to accommodate variations in the dispersion measure to the pulsar, we fit for one dispersion measure value per year (interpolating linearly between these values), plus we fit for an interplanetary medium delay (changing as our line of sight passes through different parts of the Solar System; \texttt{SOLARN0} in TEMPO2\cite{hem06}) each year.

\begin{figure}[htbp]
%\centerline{\includegraphics[width=183mm]{all-covariances.pdf}}
\centerline{\includegraphics[width=\textwidth]{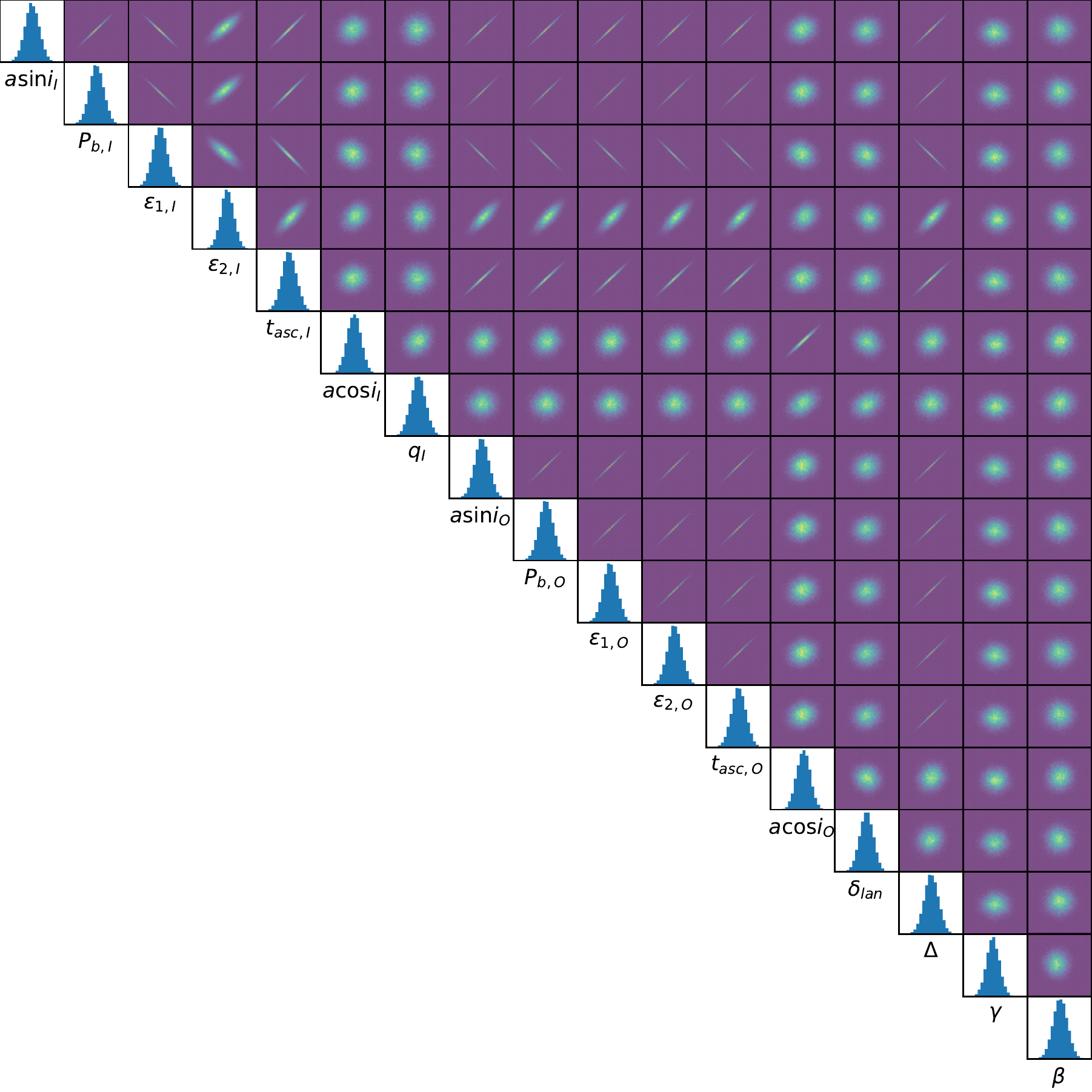}}
\caption{\label{fig:covariances}\textbf{All the covariances between parameters that affect the orbit.} This plot does not include the parameters that are evaluated by linear least-squares fitting and marginalized out. Plots on the diagonal are single-parameter histograms; plots off the diagonal are pairwise two-dimensional histograms. See Table~\ref{tab:solutionfitted} for parameter definitions.}
\end{figure}
% Delete all numbers

The reduced-dimensionality fitting problem for the non-linear parameters has some strong covariances (Figure~\ref{fig:covariances}) in spite of our attempts to choose a natural parametrization of the orbit. Nevertheless we note that the posterior distribution seems to be multivariate normal, so Bayesian methods should agree with simpler frequentist calculations.

\subsection*{Fit quality and systematics}

\begin{figure}[htbp]
\centerline{\includegraphics[width=\textwidth]{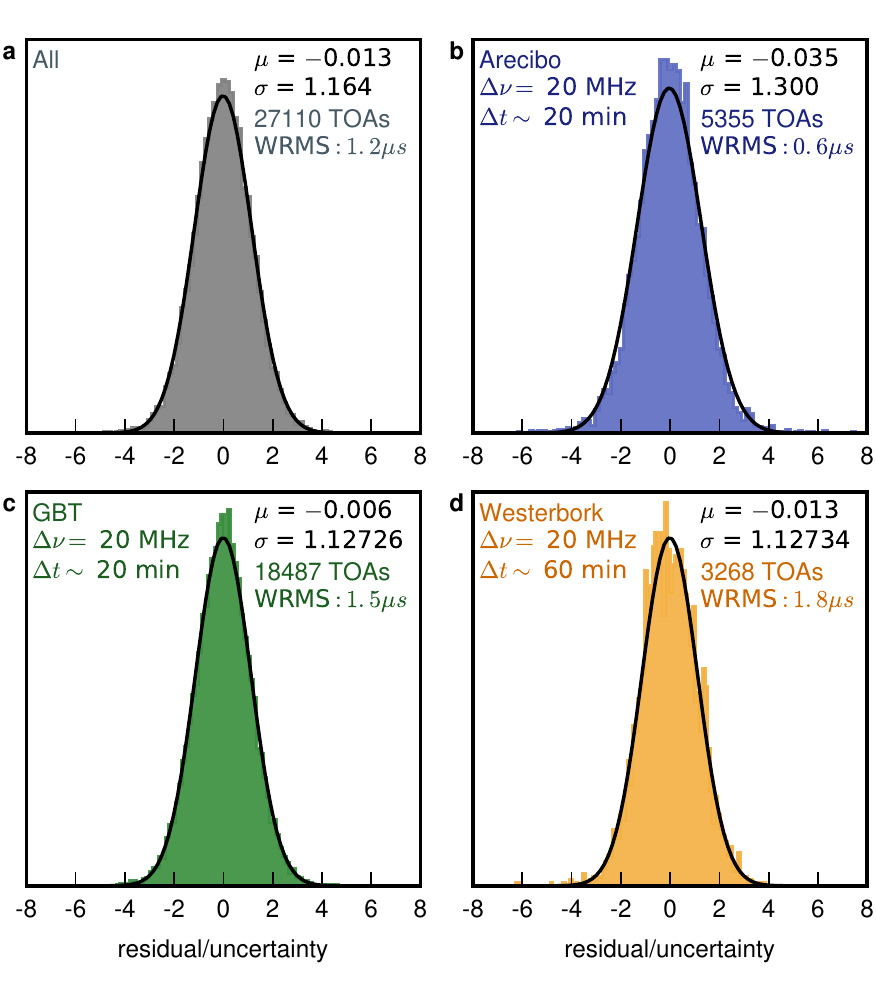}}
\caption{\label{fig:efacs}\textbf{Distribution of residual divided by uncertainty, for each telescope.} Standard deviation $\sigma$ represents the factor by which the scatter of the post-fit residuals exceeds the claimed uncertainties on pulse arrival times. $\mu$ is the mean of the distribution. Each colour represents a different telescope. Only observations in the 1400 MHz frequency band are shown here. Here $\Delta \nu$ and $\Delta t$ are the bandwidth and time, respectively, over which the data were averaged to produce each arrival time.  WRMS is the weighted root mean square of the residuals.}
\end{figure}

For a quick evaluation of fit quality, we investigated the distribution of timing residuals. Specifically, we considered the distribution of the residuals divided by the formal uncertainties on the corresponding data points (Figure~\ref{fig:efacs}). In an ideal situation this distribution should be Gaussian with zero mean and standard deviation $1$; in our data the actual scatter exceeds the claimed uncertainties by a factor of $1.164$. Although it is common practice in pulsar timing to rescale the formal uncertainties for each telescope so that $\sigma=1$, in our data there are strong correlations between residuals, which motivates a more sophisticated approach to systematics.       

Once we had selected a best-fit solution, we computed derivatives of the pulse arrival times with respect to all parameters of the solution. For the linear parameters, no additional computation was necessary, but for the parameters that affect the orbit we computed numerical derivatives using the Python package \texttt{numdifftools}. With this set of partial derivatives we could compute the ``signature'' of $\Delta$: we took the partial derivative with respect to $\Delta$ and then least-squares fit and subtracted the partial derivatives with respect to all other parameters. This produced the structure in the pulse arrival times that is uniquely explainable by a change in $\Delta$. Examining this signature, we found that it was nearly sinusoidal, with a frequency $2f_{\text{inner}}-f_{\text{outer}}$ and a specific phase. The amplitude of this sinusoid also gave us a conversion factor between amplitude in nanoseconds and size of $\Delta$ (which is dimensionless): if $\Delta$ is $10^{-6}$, then the signature will be a sinusoidal variation in pulse arrival times with amplitude $30\,\text{ns}$.

Why should the signature of an SEP violation be a sinusoid with frequency $2f_{\text{inner}}-f_{\text{outer}}$? In lunar laser ranging experiments\cite{nord68,hm18}, the signature of an SEP violation is an offset of the inner orbit in the direction of the outer companion; in Earth-Moon ranging data this should be a signal with frequency $f_{\text{inner}} - f_{\text{outer}}$, and indeed the same results are obtained when one searches for this sinusoid in the residuals from a general-relativity-obeying fit as when one introduces an SEP violation into the physics being integrated\cite{hm18}. Our observations differ in a key way from the lunar laser ranging observations: while lunar laser ranging directly measures the Earth-Moon distance, in the PSR~J0337+1715 system we measure the line-of-sight distance to the pulsar but not (directly) the distance between the pulsar and its inner companion. If, because of an SEP violation, the inner binary separation varies with a frequency $f_{\text{inner}}-f_{\text{outer}}$, and the pulsar orbits the inner centre of mass with frequency $f_{\text{inner}}$, then our line-of-sight distance measurements vary as the product of sinusoids at these two frequencies. Therefore an SEP violation should produce sinusoids at frequencies of both $2f_{\text{inner}}-f_{\text{outer}}$ and $f_{\text{outer}}$. The latter frequency can readily be absorbed into fitting the orbital parameters, which are not known a priori, but the former frequency cannot; it is the unique signature of an SEP violation.

We were concerned about systematics that affected our measurement of $\Delta$. The signature we computed above showed us what structure a systematic should have to influence $\Delta$: a sinusoid of frequency $2f_{\text{inner}}-f_{\text{outer}}$. Of course, at this particular frequency we cannot directly distinguish between systematics and genuine physical deviations from general relativity. We therefore looked at a collection of harmonically related frequencies, most of which are not associated with known physical effects related to the motion of PSR~J0337+1715. Specifically, we looked at frequencies $kf_{\text{inner}}+lf_{\text{outer}}$ for modest integer values of $k$ and $l$. Given any collection of residuals, we can compute a quasi-Fourier representation by least-squares fitting a family of sinusoids to the residuals. Each sinusoid is of the form:

\begin{equation*}
h_{k,l}(t)= {\rm I\!Re} \left( C_{k,l}\, \,  e^{2\pi i (k f_{\text{inner}} + l f_{\text{outer}})t}\right).
\label{eq:quasifourier}
\end{equation*}
This is not exactly a two-dimensional Fourier series because our data is unevenly sampled, but it resembles such a representation and the sinusoids are approximately orthogonal. Figure~\ref{fig:arrowplot} shows such a quasi-Fourier representation of the residuals from our primary fit, as well as a quasi-Fourier representation of the signature of $\Delta$. Most quasi-Fourier coefficients exceed the formal uncertainties (ellipses), indicating that there is structure in our data at these frequencies. Other quasi-Fourier coefficients, those directly related to modeled physical effects, are nearly zero because the fitting procedure has removed most or all of the power at those frequencies.  The signature of $\Delta$ (red arrow) appears almost entirely in a single quasi-Fourier coefficient, so if we can estimate the typical sizes of quasi-Fourier coefficients due to systematics, we can infer the systematic contribution to $\Delta$. 

To estimate the typical systematic contribution, we assumed all the Fourier coefficients were drawn from the same normal distribution, with mean zero. If we could compute the standard deviation of this distribution, then we would know the probability distribution of the quasi-Fourier coefficient that looks like the signature of $\Delta$; we could then infer the systematic uncertainty on our estimate of $\Delta$.

The challenge, in working with the Fourier coefficients, is that the fitting process unavoidably removes power from some of them. For example, the coefficients at $f_{\text{inner}}$ are removed by fitting for $a\sin i_I$ and $T_{\text{asc},I}$. The full nonlinear fitting process is computationally expensive and requires manual intervention, so we relied on the numerical derivatives we computed earlier to carry out a linear approximation to our fitting procedure. Thus we repeatedly generated synthetic sets of residuals with systematics drawn from a normal distribution with unit standard deviation. For each synthetic set of residuals, we (linear least-squares) fit and removed the derivatives with respect to all parameters. We then computed the power in the remaining Fourier coefficients, and scaled the synthetic data set so that this remaining power matched that in the real data set. During the fitting we also obtained the systematic contribution to $\Delta$ (appropriately rescaled). We collected these $\Delta$ values from $10^5$ synthetic data sets and then used this distribution to obtain $1\sigma$ and $95\%$ limits on $\Delta$.

\subsection*{Timing solution summary}

We have divided the system parameters into three categories: those we fixed in the fitting (Table~\ref{tab:solutionfixed}), those we fit for directly (Table~\ref{tab:solutionfitted}), and those we inferred from the fit parameters (taking covariances into account; Table~\ref{tab:solutioninferred}). 

We note that although our fit allowed astrometric parameters to vary, they have strong covariances because the ecliptic latitude of the system is only $2.1\degree$, and they are readily affected by year-long systematics such as uncorrected interplanetary medium effects. We therefore recommend against using the values quoted here for astrometric purposes. In an upcoming work we plan to compare astrometry derived from pulsar timing with that obtained from a very-long baseline interferometry campaign. 

% Please do not edit table.tex - it is automatically generated.
% This should really be in a \begin{table}, but Nature re-does all that sort of thing anyway, and there's no room for both this table and its caption in a one-page float.
% If the table grows much (e.g. astrometry fits, DMX) it may make sense to split it in two - fitted parameters and inferred parameters.
%\input{longtable.tex}
% DO NOT EDIT - automatically generated
\begin{table}
\caption{\label{tab:solutionfixed}\textbf{Fixed values and characteristics of the data set.}}
\begin{minipage}{\textwidth}
\begin{center}
\footnotesize
\begin{tabular}{lcc}
Parameter & Symbol & Value \\
\hline
Right ascension\footnote{We used the same position as in the discovery paper\cite{rsa+14}.}\footnote{Although these parameters were held fixed, we fit for offsets from these astrometric parameters; see Table~\ref{tab:solutionfitted}.} & RA & $03^h 37^m 43^s.82589$\\
Declination & Dec & $+17\degree 15' 14'' .828$\\
Parallax\footnote{We used the distance estimate, based on white dwarf modelling, given in the discovery paper\cite{rsa+14}.} & $\pi$ & $0.770$ milliarcseconds (mas)\\
Dispersion measure\footnote{We determined DM from a global fit that assumed no dependence of the profile on frequency.} & DM & $21.315933$\,pc\,cm$^{-3}$\\
Solar system ephemeris &  & DE435\\
Time conversion ephemeris &  & IF99\cite{if99}\\
Time scale &  & TCB\footnote{TCB is Barycentric Coordinate Time, a time scale that runs at a slightly different rate (faster by $1.550505\times 10^{-8}$) than terrestrial clocks because it has been corrected for the gravitational time dilation due to the Solar System potential.}\\
Reference epoch &  & MJD 55920.0\\
Observation span &  & MJD 55956.7--57866.9\\
Number of TOAs &  & 27110\\
Root-mean-squared residual (weighted) &  & $1.2\,\mu$s\\
\hline
\end{tabular}
\end{center}
\end{minipage}
\end{table}

% DO NOT EDIT - automatically generated
\begin{table}
\caption{\label{tab:solutionfitted}\textbf{Fitted values.}}
\begin{minipage}{\textwidth}
\begin{center}
\footnotesize
\begin{tabular}{lcc}
Parameter & Symbol & Value \\
\hline
\multicolumn{3}{c}{Pulsar spin parameters}\\
Pulsar spin frequency\footnote{Values in parentheses represent $1\sigma$ errors in the last decimal place(s), as determined by our MCMC fitting.}\footnote{The pulsar's proper time scale is slowed due to both gravitational time dilation and the transverse Doppler effect, in a way that varies as it moves around its orbit. Counter to the usual practice in pulsar timing, we do not correct this time scale so its average rate matches that on Earth.} & $f$ & $365.953363080(4)$ Hz\\
Pulsar spin frequency derivative & $\dot f$ & $-2.355208(3)\times 10^{ -15 }$ Hz s$^{-1}$\\
\multicolumn{3}{c}{Astrometric parameters\footnote{In order to avoid having incorrect astrometry affect our SEP test, we use derivatives to fit for offsets between fixed astrometric parameters, above, and astrometry as determined from timing. As we have not carefully analyzed systematics affecting these parameters we do not recommend further use of these timing-derived astrometric parameters.}}\\
Right ascension offset & $\Delta RA$ & $10.4(2)$ mas\\
Declination offset & $\Delta DEC$ & $-22.8(10)$ mas\\
Proper motion in right ascension & $\mu_{RA}$ & $4.51(6)$ mas/yr\\
Proper motion in declination & $\mu_{DEC}$ & $2.2(2)$ mas/yr\\
Parallax offset & $\Delta \pi$ & $-7.5(12)\times 10^{ -2 }$ mas\\
\multicolumn{3}{c}{Inner Keplerian parameters for pulsar orbit}\\
Semi-major axis projected along line of sight & $(a \sin i)_I$ & $1.2175252(2)$ lt-s\\
Orbital period & $P_{b,I}$ & $1.6293932(6)$ d\\
Eccentricity parameter\footnote{The Laplace-Lagrange parameters\cite{lk04} $\epsilon_1$ and $\epsilon_2$ provide a parameterization of the eccentricity ($e$) and longitude of periastron ($\Omega$) of an orbit that avoids a coordinate singularity at zero eccentricity; the pair $(\epsilon_2,\epsilon_1)$ forms a vector in the plane of the orbit called the eccentricity vector.} ($e\sin \omega$) & $\epsilon_{1,I}$ & $6.8833(20)\times 10^{ -4 }$ \\
Eccentricity parameter ($e\cos \omega$) & $\epsilon_{2,I}$ & $-9.1401(17)\times 10^{ -5 }$ \\
Time of ascending node & $t_{\text{asc},I}$ & MJD $55920.40771662(6)$\\
\multicolumn{3}{c}{Outer Keplerian parameters for centre of mass of inner binary}\\
Semi-major axis projected along line of sight & $(a \sin i)_O$ & $74.672629(13)$ lt-s\\
Orbital period & $P_{b,O}$ & $327.25685(11)$ d\\
Eccentricity parameter ($e\sin \omega$) & $\epsilon_{1,O}$ & $3.518595(5)\times 10^{ -2 }$ \\
Eccentricity parameter ($e\cos \omega$) & $\epsilon_{2,O}$ & $-3.46313(17)\times 10^{ -3 }$ \\
Time of ascending node & $t_{\text{asc},O}$ & MJD $56233.93512(11)$\\
\multicolumn{3}{c}{Orbital interaction parameters}\\
Semi-major axis projected in plane of sky & $(a \cos i)_I$ & $1.48950(19)$ lt-s\\
Semi-major axis projected in plane of sky & $(a \cos i)_O$ & $91.358(12)$ lt-s\\
Ratio of inner comp. mass to PSR mass & $q_I = m_{cI}/m_p$ & $0.137405(4)$ \\
Difference in longs. of asc. nodes\footnote{For a single orbit, the ascending node is the place where the pulsar passes through the plane of the sky moving away from us; the longitude of the ascending node specifies the orientation of the orbit on the sky. This is not measurable with the data we have, but the difference between the longitudes of the ascending nodes of the two orbits is measurable through orbital interactions.} & $\delta_{\text{lan}}$ & $1.2(4)\times 10^{ -4 }$ $\degree$\\
\multicolumn{3}{c}{GR violation parameters}\\
PPN nonlinearity-of-gravity parameter\footnote{The PPN parameters $\beta$ and $\gamma$ are not substantially constrained by our observations, and their values and uncertainties are consistent with the priors we obtained from Solar System experiments.} & $\beta_{\text{PPN}}-1$ & $0(3)\times 10^{ -3 }$ \\
PPN spacetime curvature parameter & $\gamma_{\text{PPN}}-1$ & $0(2)\times 10^{ -5 }$ \\
SEP violation parameter\footnote{The SEP violation parameter $\Delta$ is the fractional difference in acceleration between the pulsar and the inner white dwarf, and because it is the focus of this paper we take additional steps to estimate the impact of systematics on it.} & $\Delta$ & $(-1.1 \pm 0.7) \times 10^{-6}$\\
\hline
\end{tabular}
\end{center}
\end{minipage}
\end{table}

% DO NOT EDIT - automatically generated
\begin{table}
\caption{\label{tab:solutioninferred}\textbf{Inferred values.}}
\begin{minipage}{\textwidth}
\begin{center}
\footnotesize
\begin{tabular}{lcc}
Parameter & Symbol & Value \\
\hline
\multicolumn{3}{c}{Pulsar properties}\\
Pulsar period & $P$ & $2.73258863256(3)$ ms\\
Pulsar period derivative & $\dot P$ & $1.758643(2)\times 10^{ -20 }$ \\
Corrected pulsar period derivative\footnote{This period derivative includes corrections for the Shklovskii effect\cite{shkl70} (using the unreliable astrometry computed here) and acceleration in the Galactic potential\cite{pb17,bovy15}.} & $\dot P$ & $1.7293(16)\times 10^{ -20 }$ \\
Inferred surface dipole magnetic field\footnote{We use the standard formulae\cite{lk04} for computing $B$, $\dot E$, and $\tau$; in particular we assume a pulsar mass of $1.4M_\Sun$ and a moment of inertia of $10^{45}\;\text{g}\;\text{cm}^2$.} & $B$ & $2.2\times 10^8$ G\\
Spin-down power & $\dot E$ & $3.4\times 10^{34}$ erg s$^{-1}$\\
Characteristic age & $\tau$ & $2.5\times 10^9$ y\\
\multicolumn{3}{c}{Orbital geometry}\\
Pulsar semi-major axis (inner) & $a_I$ & $1.92379(14)$ lt-s\\
Eccentricity (inner) & $e_I$ & $6.9437(20)\times 10^{ -4 }$ \\
Longitude of periastron (inner) & $\omega_I$ & $97.5638(14)$ $\degree$\\
Pulsar semi-major axis (outer) & $a_O$ & $117.992(9)$ lt-s\\
Eccentricity (outer) & $e_O$ & $3.535596(3)\times 10^{ -2 }$ \\
Longitude of periastron (outer) & $\omega_O$ & $95.6212(3)$ $\degree$\\
Inclination of invariant plane\footnote{The invariant plane is the plane perpendicular to the total (orbital) angular momentum of the triple system.} & $i$ & $39.262(4)$ $\degree$\\
Inclination of inner orbit & $i_I$ & $39.263(4)$ $\degree$\\
Angle between orbital planes & $\delta_i$ & $1.4(6)\times 10^{ -3 }$ $\degree$\\
Angle between eccentricity vectors & $\delta_\omega \sim \omega_O-\omega_I$ & $-1.9427(16)$ $\degree$\\
Relativistic periastron advance (inner) & $\dot\omega_I$ & $0.122293(18)$ $\degree/\text{year}$\\
Relativistic periastron advance (outer) & $\dot\omega_O$ & $2.0636(3)\times 10^{ -5 }$ $\degree/\text{year}$\\
\multicolumn{3}{c}{Masses\footnote{Masses are as measured by a distant observer; corrections for special relativity are at the $10^{-8}$ fractional level and are therefore irrelevant.}}\\
Pulsar mass & $m_p$ & $1.4359(3)$ $M_\Sun$\\
Inner companion mass & $m_{cI}$ & $0.19730(4)$ $M_\Sun$\\
Outer companion mass & $m_{cO}$ & $0.40962(9)$ $M_\Sun$\\
\hline
\end{tabular}
\end{center}
\end{minipage}
\end{table}

% table testing
%\clearpage
%\begin{table}
%\caption{\textbf{A toy table experiment.}}
%\begin{minipage}{\textwidth}
%\begin{center}
%\begin{tabular}{lcr}
%one & two\footnote{two is the smallest prime} & three\footnote{three is prime too} \\
%\hline
%four & five & six \\
%\end{tabular}
%\end{center}
%\end{minipage}
%\end{table}

\subsection*{Orbital effects}

In our timing model we used first-order post-Newtonian (1PN) equations of motion of three point particles, i.e. we neglected orbital effects caused by tidal deformation of the stars as well as higher-order post-Newtonian effects such as frame dragging\cite{thir18} and gravitational wave emission\cite{eins16b}. These effects strongly depend on the distance between objects, and, given that even the inner binary of the triple system has a relatively wide orbit (16 light-seconds), they are too small to affect our data. Here we estimate the impact of these effects on the measured orbital parameters. 

\emph{Periastron advance:} The first post-Newtonian order relativistic periastron advance and classical periastron advance caused by three-body interactions are included in the fit, since they are taken into account in the equations of motion. Some additional periastron advance can be caused by tidal deformation of the inner white dwarf companion of the system\cite{russ28}, or by any higher-order post-Newtonian effects\cite{eins16,thir18}. We calculated the change of the longitude of periastron of the pulsar orbit caused by tidal asymmetry of the companion\cite{sb76,gu+17} assuming a tidal Love number $k_2\sim 0.01$, appropriate for a helium-core white dwarf with an extended envelope\cite{pm12}: $\Delta {\omega}_{tidal} = 3 \times 10^{-5}$$\, \degree/\text{year}$. This value is about five orders of magnitude smaller than relativistic periastron advance\cite{eins16} $\Delta {\omega}_{rel}=0.12\degree/\text{year}$ and about three orders of magnitude smaller than what can be detected with the current precision of the existing data (see Table~\ref{tab:solutionfitted}). At higher post-Newtonian order the periastron advance of the inner orbit is dominated by the Lense-Thirring effect, where interaction of the orbital angular momentum and the spin of the inner white dwarf cause the orbital plane to precess\cite{Wex14}. We estimate this precession to be no greater than $\Omega_{SO}=3 \times 10^{-4}\, \degree/\text{year}$, even if the white dwarf is rotating as rapidly as once per minute and its angular momentum is aligned with the angular momentum of the inner orbit.

\emph{Dissipative effects:} The tidal deformation of the inner white dwarf can cause a loss of energy from the system (tidal lag), thereby shrinking the orbit. Assuming that the effective tidal parameter $Q \approx 10^{7}$ for the inner white dwarf\cite{pm12,lg14}, we calculated the characteristic time scale of the orbital period decrease\cite{sb76} due to this effect: $\tau_p \equiv P_{orb}/\dot{P}_{orb}\approx 2 \times 10^{17} \, \mathrm{yr}$. The characteristic timescale of the orbital decay due to gravitational wave emission $\tau_{gr} \approx 2 \times 10^{12}$ yr. The current precision with which we can measure $P_{orb}$ of the inner binary is $\approx 10^{-8}$ days. This means that we need about $100$ years of observations in order to be able to detect dissipation of the inner binary orbit due to gravitational wave emission and about $10^{7}$ years to detect the dissipation due to tidal deceleration.

\subsection*{Theoretical implications}

Our central result is that in the same gravitational field, the fractional difference in accelerations $|\Delta|$ between a 1.4 $M_\Sun$ pulsar and a white dwarf is no more than $2.6\times 10^{-6}$ (at $95\%$ confidence). We can directly compare this to a previous test in which the pulsar PSR~J1713+0747 and its white dwarf companion were observed falling in the Galactic potential\cite{zdw+18}; this test constrains $|\Delta| < 2\times 10^{-3}$ ($95\%$), so our result is an improvement by three orders of magnitude. That said, we would like to compare our result to the weak-field SEP test carried out by lunar laser ranging, or to indirect limits on SEP violations coming from upper bounds on gravitational dipole radiation. Such comparisons require a theoretical framework.

Lunar laser ranging places a limit $|\Delta|<1.3 \times 10^{-14}$ on the Earth-Moon-Sun system\cite{hm18}. Since this system is well-approximated by the first post-Newtonian order, we can describe SEP violations by the (weak-field) Nordtvedt parameter $\eta_N$: $\Delta = \eta_N E_B$, where $E_B$ is the fractional gravitational binding energy of the test body. Since $E_B$ for the earth is $-4.45\times 10^{-10}$, lunar laser ranging constrains\cite{hm18} $|\eta_N|<2.4\times 10^{-4}$ ($95\%$). If we simply apply this to the PSR~J0337+1715 system, $E_B \sim 0.1$ and $|\Delta|<2.6\times 10^{-6}$, so the (strong-field) Nordtvedt parameter $|\hat\eta_N| < 2.6\times 10^{-5}$, and we appear to dramatically improve upon the lunar laser ranging result. However, the phenomenon of (potential) SEP violation arises from the interior of the pulsar, where the first post-Newtonian order is an insufficient approximation. Directly comparing weak-field $\eta_N$ and strong-field $\hat\eta_N$ depends on having a strong-field theory of gravity. Unfortunately, no completely general framework exists for describing strong-field effects, so it is necessary to specialize somewhat to specific families of strong-field gravity theories. Berti et al.\cite{bbc+15} provide an overview of the space of possibilities.

Within the context of tensor-(multi-)scalar theories, there is a parametrization of the second post-Newtonian (2PN) order in terms of only four parameters\cite{de96b}: the $\beta$ and $\gamma$ of standard parametrized post-Newtonian models, and $\epsilon$ and $\zeta$, which describe the 2PN effects. Combining the lunar laser ranging constraint on $\eta_N = 4\beta - \gamma -1$ with our constraint on $\Delta$, we can infer a limit of $|\epsilon/2+\zeta|<10^{-3}$. Nevertheless, in the interior of a neutron star, the 2PN approximation may not be sufficient either.

Horbatsch and Burgess suggest\cite{hb12} describing pulsar timing results in terms of constraints on scalar coupling constants $\alpha_j$ for the bodies involved. The values for these coupling constants depend on the scalar-tensor theory being considered and the equation of state assumed for the neutron star. Expressing the constraints in this way gathers almost all the theory and equation of state dependence in these $\alpha_j$, providing a somewhat theory-independent way to compare pulsar-timing results. In our case, this is quite straightforward, as $|\Delta| = |\alpha_o(\alpha_p-\alpha_i)| < 2.6\times 10^{-6}$. Here $\alpha_i$ and $\alpha_o$ are coupling constants for the inner and outer white dwarfs respectively, equal to a weak-field coupling constant in many theories, while $\alpha_p$ is the scalar coupling constant for this pulsar, which has mass $1.4359(3) M_\Sun$. Limits on dipole gravitational radiation constrain the combination $(\alpha_p - \alpha_i)^2$, implying a somewhat different dependence on theory parameters. Some theories, including TeVeS\cite{beke04}, predict that $\alpha_p = \alpha_i = \alpha_o$ and therefore these theories cannot be constrained by our result\cite{sw16}.
% double-check specific number

Taylor et al.\cite{twdw92} introduce a family of tensor-multi-scalar theories whose first post-Newtonian order terms agree exactly with general relativity but which have strong-field behaviour parametrized by $\beta'$ and $\beta''$. Our $\Delta$ measurement implies $|\beta'|<3.5\times 10^{-3}$; combining this with existing results\cite{twdw92} implies also $|\beta''|<1$. This family of theories, unfortunately, suffers from some serious theoretical problems, including the presence of negative-energy excitations.
% Cite negative-energy

The standard framework for comparing strong-field tests of general relativity is the quasi-Brans-Dicke theories, parametrized by $\alpha_0$ and $\beta_0$. Standard Brans-Dicke gravity arises in the special case $\beta_0=0$, where the Brans-Dicke parameter $\omega_{BD}$ is related to $\alpha_0$ by $\alpha_0^2 = (2\omega_{BD}+3)^{-1}$. 
Quasi-Brans-Dicke theories can, under certain circumstances, arise as local approximations to theories in which the scalar has a potential that causes it to vary on cosmological scales\cite{de96}. In Solar System tests, the relevant quantity is $|\alpha_0|$, and by making this small enough weak-field deviations from general relativity can be suppressed enough to pass any given test. In standard Brans-Dicke gravity, the Cassini Shapiro delay measurement\cite{bit03} limits $\omega_BD$ to be greater than about $15000$; the constraint we derive implies $\omega_BD$ is greater than about $73000$ ($95\%$). However, in quasi-Brans-Dicke gravity, for sufficiently negative values of $\beta_0$ massive pulsars can undergo ``spontaneous scalarization'' and acquire an order-unity deviation from general relativity regardless of the theory's weak-field behaviour. In other words, regardless of how well Solar System tests constrain the weak-field behaviour of gravity, this family of theories can still exhibit strong-field behaviour that differs substantially from general relativity.

Using this quasi-Brans-Dicke family of theories, combined with a neutron star equation of state, we can compare SEP tests with pulsars of different masses and in the weak field, and we can also compare our result with tests based on the absence of dipole gravitational radiation. Dipole gravitational radiation can only arise if the centre of gravitational mass of a binary is not the same as the centre of inertial mass. Thus dipole gravitational radiation implies an SEP violation, but relating such upper limits to direct SEP tests requires a specific theory. Within the quasi-Brans-Dicke framework, if we choose an equation of state, we can integrate the generalized Tolman-Oppenheimer-Volkoff equations\cite{de96} to compute both $\Delta$ for that neutron star and the degree to which it produces gravitational dipole radiation. Following Antoniadis et al.\cite{afw+13} we have chosen a very stiff equation of state (``.20'' of Haensel et al.\cite{hpk81}; the maximum mass for a neutron star in this equation of state is $2.6 M_\Sun$; stiffer equations of state lead to less-constraining limits from pulsar-based tests). In this family of theories, and with this equation of state, we evaluated constraints: from our result, from existing wide binaries\cite{sfl+05}, from lunar laser ranging\cite{hm18}, from an SEP test with Messenger\cite{gmg+18}, from the Cassini Shapiro delay measurement\cite{bit03}, and from gravitational dipole radiation upper limits on two pulsar-white dwarf systems\cite{fwe+12,afw+13}. The results are plotted in Figure~\ref{fig:alpha-beta}. Note that the parametrized post-Newtonian parameters $\beta$ and $\gamma$ are proportional to $\alpha_0^2$, so comparing the parametrized post-Newtonian values doubles the number of orders of magnitude difference between tests as opposed to comparing $\alpha_0$. We see that the constraint derived from the motion of PSR~J0337+1715, in addition to being a direct strong-field test of the SEP, substantially improves upon existing theory constraints for most values of $\beta_0$.

%\subsection*{Code availability}
%
%All general-purpose software packages we use are open-source:
%\begin{description}
%\item[psrchive:] \url{http://psrchive.sourceforge.net/}
%\item[TEMPO:] \url{http://tempo.sourceforge.net/}
%\item[TEMPO2:] \url{http://www.atnf.csiro.au/research/pulsar/tempo2/}
%\item[numdifftools:] \url{https://pypi.python.org/pypi/Numdifftools}
%\item[sympy:] \url{http://www.sympy.org/}
%\item[boost:] \url{http://www.boost.org/}
%\end{description}

\end{methods}

%% Put the bibliography here, most people will use BiBTeX in
%% which case the environment below should be replaced with
%% the \bibliography{} command.

\bibliography{refs.bib}

\begin{thebibliography}{10}
\expandafter\ifx\csname url\endcsname\relax
  \def\url#1{\texttt{#1}}\fi
\expandafter\ifx\csname urlprefix\endcsname\relax\def\urlprefix{URL }\fi
\providecommand{\bibinfo}[2]{#2}
\providecommand{\eprint}[2][]{\url{#2}}

\bibitem{eins16}
\bibinfo{author}{{Einstein}, A.}
\newblock \bibinfo{title}{{Die Grundlage der allgemeinen
  Relativit{\"a}tstheorie}}.
\newblock \emph{\bibinfo{journal}{Annalen der Physik}}
  \textbf{\bibinfo{volume}{354}}, \bibinfo{pages}{769--822}
  (\bibinfo{year}{1916}).

\bibitem{will14}
\bibinfo{author}{{Will}, C.~M.}
\newblock \bibinfo{title}{{The Confrontation between General Relativity and
  Experiment}}.
\newblock \emph{\bibinfo{journal}{Living Reviews in Relativity}}
  \textbf{\bibinfo{volume}{17}}, \bibinfo{pages}{4} (\bibinfo{year}{2014}).
\newblock \eprint{1403.7377}.

\bibitem{deru11}
\bibinfo{author}{{Deruelle}, N.}
\newblock \bibinfo{title}{{Nordstr{\"o}m's scalar theory of gravity and the
  equivalence principle}}.
\newblock \emph{\bibinfo{journal}{General Relativity and Gravitation}}
  \textbf{\bibinfo{volume}{43}}, \bibinfo{pages}{3337--3354}
  (\bibinfo{year}{2011}).
\newblock \eprint{1104.4608}.

\bibitem{dp94}
\bibinfo{author}{{Damour}, T.} \& \bibinfo{author}{{Polyakov}, A.~M.}
\newblock \bibinfo{title}{{The string dilation and a least coupling
  principle}}.
\newblock \emph{\bibinfo{journal}{Nuclear Physics B}}
  \textbf{\bibinfo{volume}{423}}, \bibinfo{pages}{532--558}
  (\bibinfo{year}{1994}).
\newblock \eprint{hep-th/9401069}.

\bibitem{hm18}
\bibinfo{author}{{Hofmann}, F.} \& \bibinfo{author}{{M{\"u}ller}, J.}
\newblock \bibinfo{title}{{Relativistic tests with lunar laser ranging}}.
\newblock \emph{\bibinfo{journal}{Classical and Quantum Gravity}}
  \textbf{\bibinfo{volume}{35}}, \bibinfo{pages}{035015}
  (\bibinfo{year}{2018}).

\bibitem{zdw+18}
\bibinfo{author}{{Zhu}, W.~W.} \emph{et~al.}
\newblock \bibinfo{title}{{Tests of Gravitational Symmetries with Pulsar Binary
  J1713+0747}}.
\newblock \emph{\bibinfo{journal}{ArXiv e-prints}}  (\bibinfo{year}{2018}).
\newblock \eprint{1802.09206}.

\bibitem{rsa+14}
\bibinfo{author}{{Ransom}, S.~M.} \emph{et~al.}
\newblock \bibinfo{title}{{A millisecond pulsar in a stellar triple system}}.
\newblock \emph{\bibinfo{journal}{\nat}} \textbf{\bibinfo{volume}{505}},
  \bibinfo{pages}{520--524} (\bibinfo{year}{2014}).
\newblock \eprint{1401.0535}.

\bibitem{gmg+18}
\bibinfo{author}{{Genova}, A.} \emph{et~al.}
\newblock \bibinfo{title}{{Solar system expansion and strong equivalence
  principle as seen by the NASA MESSENGER mission}}.
\newblock \emph{\bibinfo{journal}{Nature Communications}}
  \textbf{\bibinfo{volume}{9}}, \bibinfo{pages}{289} (\bibinfo{year}{2018}).

\bibitem{nab+15}
\bibinfo{author}{{The NANOGrav Collaboration}} \emph{et~al.}
\newblock \bibinfo{title}{{The NANOGrav Nine-year Data Set: Observations,
  Arrival Time Measurements, and Analysis of 37 Millisecond Pulsars}}.
\newblock \emph{\bibinfo{journal}{\apj}} \textbf{\bibinfo{volume}{813}},
  \bibinfo{pages}{65} (\bibinfo{year}{2015}).
\newblock \eprint{1505.07540}.

\bibitem{wn72}
\bibinfo{author}{{Will}, C.~M.} \& \bibinfo{author}{{Nordtvedt}, K., Jr.}
\newblock \bibinfo{title}{{Conservation Laws and Preferred Frames in
  Relativistic Gravity. I. Preferred-Frame Theories and an Extended PPN
  Formalism}}.
\newblock \emph{\bibinfo{journal}{\apj}} \textbf{\bibinfo{volume}{177}},
  \bibinfo{pages}{757} (\bibinfo{year}{1972}).

\bibitem{nord85}
\bibinfo{author}{{Nordtvedt}, K.}
\newblock \bibinfo{title}{{A post-Newtonian gravitational Lagrangian formalism
  for celestial body dynamics in metric gravity}}.
\newblock \emph{\bibinfo{journal}{\apj}} \textbf{\bibinfo{volume}{297}},
  \bibinfo{pages}{390--404} (\bibinfo{year}{1985}).

\bibitem{msp+17}
\bibinfo{author}{Meurer, A.} \emph{et~al.}
\newblock \bibinfo{title}{Sympy: symbolic computing in python}.
\newblock \emph{\bibinfo{journal}{PeerJ Computer Science}}
  \textbf{\bibinfo{volume}{3}}, \bibinfo{pages}{e103} (\bibinfo{year}{2017}).
\newblock \urlprefix\url{https://doi.org/10.7717/peerj-cs.103}.

\bibitem{ds91}
\bibinfo{author}{{Damour}, T.} \& \bibinfo{author}{{Schaefer}, G.}
\newblock \bibinfo{title}{{New tests of the strong equivalence principle using
  binary-pulsar data}}.
\newblock \emph{\bibinfo{journal}{Physical Review Letters}}
  \textbf{\bibinfo{volume}{66}}, \bibinfo{pages}{2549--2552}
  (\bibinfo{year}{1991}).

\bibitem{bit03}
\bibinfo{author}{{Bertotti}, B.}, \bibinfo{author}{{Iess}, L.} \&
  \bibinfo{author}{{Tortora}, P.}
\newblock \bibinfo{title}{{A test of general relativity using radio links with
  the Cassini spacecraft}}.
\newblock \emph{\bibinfo{journal}{\nat}} \textbf{\bibinfo{volume}{425}},
  \bibinfo{pages}{374--376} (\bibinfo{year}{2003}).

\bibitem{bps15}
\bibinfo{author}{{Baker}, T.}, \bibinfo{author}{{Psaltis}, D.} \&
  \bibinfo{author}{{Skordis}, C.}
\newblock \bibinfo{title}{{Linking Tests of Gravity on All Scales: from the
  Strong-field Regime to Cosmology}}.
\newblock \emph{\bibinfo{journal}{\apj}} \textbf{\bibinfo{volume}{802}},
  \bibinfo{pages}{63} (\bibinfo{year}{2015}).
\newblock \eprint{1412.3455}.

\bibitem{sfl+05}
\bibinfo{author}{{Stairs}, I.~H.} \emph{et~al.}
\newblock \bibinfo{title}{{Discovery of Three Wide-Orbit Binary Pulsars:
  Implications for Binary Evolution and Equivalence Principles}}.
\newblock \emph{\bibinfo{journal}{\apj}} \textbf{\bibinfo{volume}{632}},
  \bibinfo{pages}{1060--1068} (\bibinfo{year}{2005}).
\newblock \eprint{astro-ph/0506188}.

\bibitem{fwe+12}
\bibinfo{author}{{Freire}, P.~C.~C.} \emph{et~al.}
\newblock \bibinfo{title}{{The relativistic pulsar-white dwarf binary PSR
  J1738+0333 - II. The most stringent test of scalar-tensor gravity}}.
\newblock \emph{\bibinfo{journal}{\mnras}} \textbf{\bibinfo{volume}{423}},
  \bibinfo{pages}{3328--3343} (\bibinfo{year}{2012}).
\newblock \eprint{1205.1450}.

\bibitem{afw+13}
\bibinfo{author}{{Antoniadis}, J.} \emph{et~al.}
\newblock \bibinfo{title}{{A Massive Pulsar in a Compact Relativistic Binary}}.
\newblock \emph{\bibinfo{journal}{Science}} \textbf{\bibinfo{volume}{340}},
  \bibinfo{pages}{448} (\bibinfo{year}{2013}).
\newblock \eprint{1304.6875}.

\bibitem{bbc+15}
\bibinfo{author}{{Berti}, E.} \emph{et~al.}
\newblock \bibinfo{title}{{Testing general relativity with present and future
  astrophysical observations}}.
\newblock \emph{\bibinfo{journal}{Classical and Quantum Gravity}}
  \textbf{\bibinfo{volume}{32}}, \bibinfo{pages}{243001}
  (\bibinfo{year}{2015}).
\newblock \eprint{1501.07274}.

\bibitem{bd61}
\bibinfo{author}{{Brans}, C.} \& \bibinfo{author}{{Dicke}, R.~H.}
\newblock \bibinfo{title}{{Mach's Principle and a Relativistic Theory of
  Gravitation}}.
\newblock \emph{\bibinfo{journal}{Physical Review}}
  \textbf{\bibinfo{volume}{124}}, \bibinfo{pages}{925--935}
  (\bibinfo{year}{1961}).

\bibitem{de92}
\bibinfo{author}{{Damour}, T.} \& \bibinfo{author}{{Esposito-Farese}, G.}
\newblock \bibinfo{title}{{Tensor-multi-scalar theories of gravitation}}.
\newblock \emph{\bibinfo{journal}{Classical and Quantum Gravity}}
  \textbf{\bibinfo{volume}{9}}, \bibinfo{pages}{2093--2176}
  (\bibinfo{year}{1992}).

\bibitem{drd+08}
\bibinfo{author}{{DuPlain}, R.} \emph{et~al.}
\newblock \bibinfo{title}{{Launching GUPPI: the Green Bank Ultimate Pulsar
  Processing Instrument}}.
\newblock In \emph{\bibinfo{booktitle}{Advanced Software and Control for
  Astronomy II}}, vol. \bibinfo{volume}{7019} of
  \emph{\bibinfo{series}{\procspie}}, \bibinfo{pages}{70191D}
  (\bibinfo{year}{2008}).

\bibitem{kss08}
\bibinfo{author}{{Karuppusamy}, R.}, \bibinfo{author}{{Stappers}, B.} \&
  \bibinfo{author}{{van Straten}, W.}
\newblock \bibinfo{title}{{PuMa-II: A Wide Band Pulsar Machine for the
  Westerbork Synthesis Radio Telescope}}.
\newblock \emph{\bibinfo{journal}{\pasp}} \textbf{\bibinfo{volume}{120}},
  \bibinfo{pages}{191} (\bibinfo{year}{2008}).
\newblock \eprint{0802.2245}.

\bibitem{hr75}
\bibinfo{author}{{Hankins}, T.~H.} \& \bibinfo{author}{{Rickett}, B.~J.}
\newblock \bibinfo{title}{{Pulsar signal processing}}.
\newblock In \bibinfo{editor}{{Alder}, B.}, \bibinfo{editor}{{Fernbach}, S.} \&
  \bibinfo{editor}{{Rotenberg}, M.} (eds.) \emph{\bibinfo{booktitle}{Methods in
  Computational Physics. Volume 14 - Radio astronomy}},
  vol.~\bibinfo{volume}{14}, \bibinfo{pages}{55--129} (\bibinfo{year}{1975}).

\bibitem{stra06}
\bibinfo{author}{{van Straten}, W.}
\newblock \bibinfo{title}{{Radio Astronomical Polarimetry and High-Precision
  Pulsar Timing}}.
\newblock \emph{\bibinfo{journal}{\apj}} \textbf{\bibinfo{volume}{642}},
  \bibinfo{pages}{1004--1011} (\bibinfo{year}{2006}).
\newblock \eprint{astro-ph/0510334}.

\bibitem{lr99}
\bibinfo{author}{{Lambert}, H.~C.} \& \bibinfo{author}{{Rickett}, B.~J.}
\newblock \bibinfo{title}{{On the Theory of Pulse Propagation and Two-Frequency
  Field Statistics in Irregular Interstellar Plasmas}}.
\newblock \emph{\bibinfo{journal}{\apj}} \textbf{\bibinfo{volume}{517}},
  \bibinfo{pages}{299--317} (\bibinfo{year}{1999}).

\bibitem{akhs14}
\bibinfo{author}{{Archibald}, A.~M.}, \bibinfo{author}{{Kondratiev}, V.~I.},
  \bibinfo{author}{{Hessels}, J.~W.~T.} \& \bibinfo{author}{{Stinebring},
  D.~R.}
\newblock \bibinfo{title}{{Millisecond Pulsar Scintillation Studies with LOFAR:
  Initial Results}}.
\newblock \emph{\bibinfo{journal}{\apjl}} \textbf{\bibinfo{volume}{790}},
  \bibinfo{pages}{L22} (\bibinfo{year}{2014}).
\newblock \eprint{1407.0171}.

\bibitem{bs66}
\bibinfo{author}{Bulirsch, R.} \& \bibinfo{author}{Stoer, J.}
\newblock \bibinfo{title}{Asymptotic upper and lower bounds for results of
  extrapolation methods}.
\newblock \emph{\bibinfo{journal}{Numerische Mathematik}}
  \textbf{\bibinfo{volume}{8}}, \bibinfo{pages}{93--104}
  (\bibinfo{year}{1966}).
\newblock \urlprefix\url{https://doi.org/10.1007/BF02163179}.

\bibitem{hem06}
\bibinfo{author}{{Hobbs}, G.~B.}, \bibinfo{author}{{Edwards}, R.~T.} \&
  \bibinfo{author}{{Manchester}, R.~N.}
\newblock \bibinfo{title}{{TEMPO2, a new pulsar-timing package - I. An
  overview}}.
\newblock \emph{\bibinfo{journal}{\mnras}} \textbf{\bibinfo{volume}{369}},
  \bibinfo{pages}{655--672} (\bibinfo{year}{2006}).
\newblock \eprint{astro-ph/0603381}.

\bibitem{nord68}
\bibinfo{author}{{Nordtvedt}, K.}
\newblock \bibinfo{title}{{Testing Relativity with Laser Ranging to the Moon}}.
\newblock \emph{\bibinfo{journal}{Physical Review}}
  \textbf{\bibinfo{volume}{170}}, \bibinfo{pages}{1186--1187}
  (\bibinfo{year}{1968}).

\bibitem{if99}
\bibinfo{author}{{Irwin}, A.~W.} \& \bibinfo{author}{{Fukushima}, T.}
\newblock \bibinfo{title}{{A numerical time ephemeris of the Earth}}.
\newblock \emph{\bibinfo{journal}{\aap}} \textbf{\bibinfo{volume}{348}},
  \bibinfo{pages}{642--652} (\bibinfo{year}{1999}).

\bibitem{lk04}
\bibinfo{author}{{Lorimer}, D.~R.} \& \bibinfo{author}{{Kramer}, M.}
\newblock \emph{\bibinfo{title}{{Handbook of Pulsar Astronomy}}}
  (\bibinfo{year}{2004}).

\bibitem{shkl70}
\bibinfo{author}{{Shklovskii}, I.~S.}
\newblock \bibinfo{title}{{Possible Causes of the Secular Increase in Pulsar
  Periods.}}
\newblock \emph{\bibinfo{journal}{\sovast}} \textbf{\bibinfo{volume}{13}},
  \bibinfo{pages}{562} (\bibinfo{year}{1970}).

\bibitem{pb17}
\bibinfo{author}{{Pathak}, D.} \& \bibinfo{author}{{Bagchi}, M.}
\newblock \bibinfo{title}{{GalDynPsr: A package to estimate dynamical
  contributions in the rate of change of the period of radio pulsars}}.
\newblock \emph{\bibinfo{journal}{ArXiv e-prints}}  (\bibinfo{year}{2017}).
\newblock \eprint{1712.06590}.

\bibitem{bovy15}
\bibinfo{author}{{Bovy}, J.}
\newblock \bibinfo{title}{{galpy: A python Library for Galactic Dynamics}}.
\newblock \emph{\bibinfo{journal}{\apjs}} \textbf{\bibinfo{volume}{216}},
  \bibinfo{pages}{29} (\bibinfo{year}{2015}).
\newblock \eprint{1412.3451}.

\bibitem{thir18}
\bibinfo{author}{{Thirring}, H.}
\newblock \bibinfo{title}{{{\"U}ber die Wirkung rotierender ferner Massen in
  der Einsteinschen Gravitationstheorie.}}
\newblock \emph{\bibinfo{journal}{Physikalische Zeitschrift}}
  \textbf{\bibinfo{volume}{19}} (\bibinfo{year}{1918}).

\bibitem{eins16b}
\bibinfo{author}{{Einstein}, A.}
\newblock \bibinfo{title}{{N{\"a}herungsweise Integration der Feldgleichungen
  der Gravitation}}.
\newblock \emph{\bibinfo{journal}{Sitzungsberichte der K{\"o}niglich
  Preu{\ss}ischen Akademie der Wissenschaften (Berlin), Seite 688-696.}}
  (\bibinfo{year}{1916}).

\bibitem{russ28}
\bibinfo{author}{{Russell}, H.~N.}
\newblock \bibinfo{title}{{On the advance of periastron in eclipsing
  binaries}}.
\newblock \emph{\bibinfo{journal}{\mnras}} \textbf{\bibinfo{volume}{88}},
  \bibinfo{pages}{641--643} (\bibinfo{year}{1928}).

\bibitem{sb76}
\bibinfo{author}{{Smarr}, L.~L.} \& \bibinfo{author}{{Blandford}, R.}
\newblock \bibinfo{title}{{The binary pulsar - Physical processes, possible
  companions, and evolutionary histories}}.
\newblock \emph{\bibinfo{journal}{\apj}} \textbf{\bibinfo{volume}{207}},
  \bibinfo{pages}{574--588} (\bibinfo{year}{1976}).

\bibitem{gu+17}
\bibinfo{author}{Gusinskaia, N.~V.} \emph{et~al.}
\newblock \bibinfo{title}{Conquering systematics in the timing of the pulsar
  triple system j0337+1715: Towards a unique and robust test of the strong
  equivalence principle}.
\newblock \emph{\bibinfo{journal}{Journal of Physics: Conference Series}}
  \textbf{\bibinfo{volume}{932}}, \bibinfo{pages}{012003}
  (\bibinfo{year}{2017}).
\newblock \urlprefix\url{http://stacks.iop.org/1742-6596/932/i=1/a=012003}.

\bibitem{pm12}
\bibinfo{author}{{Prodan}, S.} \& \bibinfo{author}{{Murray}, N.}
\newblock \bibinfo{title}{{On the Dynamics and Tidal Dissipation Rate of the
  White Dwarf in 4U 1820-30}}.
\newblock \emph{\bibinfo{journal}{\apj}} \textbf{\bibinfo{volume}{747}},
  \bibinfo{pages}{4} (\bibinfo{year}{2012}).
\newblock \eprint{1110.6655}.

\bibitem{Wex14}
\bibinfo{author}{{Wex}, N.}
\newblock \bibinfo{title}{{Testing Relativistic Gravity with Radio Pulsars}}.
\newblock \emph{\bibinfo{journal}{ArXiv e-prints}}  (\bibinfo{year}{2014}).
\newblock \eprint{1402.5594}.

\bibitem{lg14}
\bibinfo{author}{{Luan}, J.} \& \bibinfo{author}{{Goldreich}, P.}
\newblock \bibinfo{title}{{Secular Evolution of the Pulsar Triple System
  J0337+1715}}.
\newblock \emph{\bibinfo{journal}{\apj}} \textbf{\bibinfo{volume}{790}},
  \bibinfo{pages}{82} (\bibinfo{year}{2014}).
\newblock \eprint{1405.2374}.

\bibitem{de96b}
\bibinfo{author}{{Damour}, T.} \& \bibinfo{author}{{Esposito-Far{\`e}se}, G.}
\newblock \bibinfo{title}{{Testing gravity to second post-Newtonian order: A
  field-theory approach}}.
\newblock \emph{\bibinfo{journal}{\prd}} \textbf{\bibinfo{volume}{53}},
  \bibinfo{pages}{5541--5578} (\bibinfo{year}{1996}).
\newblock \eprint{gr-qc/9506063}.

\bibitem{hb12}
\bibinfo{author}{{Horbatsch}, M.~W.} \& \bibinfo{author}{{Burgess}, C.~P.}
\newblock \bibinfo{title}{{Model-independent comparisons of pulsar timings to
  scalar-tensor gravity}}.
\newblock \emph{\bibinfo{journal}{Classical and Quantum Gravity}}
  \textbf{\bibinfo{volume}{29}}, \bibinfo{pages}{245004}
  (\bibinfo{year}{2012}).
\newblock \eprint{1107.3585}.

\bibitem{beke04}
\bibinfo{author}{{Bekenstein}, J.~D.}
\newblock \bibinfo{title}{{Relativistic gravitation theory for the modified
  Newtonian dynamics paradigm}}.
\newblock \emph{\bibinfo{journal}{\prd}} \textbf{\bibinfo{volume}{70}},
  \bibinfo{pages}{083509} (\bibinfo{year}{2004}).
\newblock \eprint{astro-ph/0403694}.

\bibitem{sw16}
\bibinfo{author}{{Shao}, L.} \& \bibinfo{author}{{Wex}, N.}
\newblock \bibinfo{title}{{Tests of gravitational symmetries with radio
  pulsars}}.
\newblock \emph{\bibinfo{journal}{Science China Physics, Mechanics, and
  Astronomy}} \textbf{\bibinfo{volume}{59}}, \bibinfo{pages}{699501}
  (\bibinfo{year}{2016}).
\newblock \eprint{1604.03662}.

\bibitem{twdw92}
\bibinfo{author}{{Taylor}, J.~H.}, \bibinfo{author}{{Wolszczan}, A.},
  \bibinfo{author}{{Damour}, T.} \& \bibinfo{author}{{Weisberg}, J.~M.}
\newblock \bibinfo{title}{{Experimental constraints on strong-field
  relativistic gravity}}.
\newblock \emph{\bibinfo{journal}{\nat}} \textbf{\bibinfo{volume}{355}},
  \bibinfo{pages}{132--136} (\bibinfo{year}{1992}).

\bibitem{de96}
\bibinfo{author}{{Damour}, T.} \& \bibinfo{author}{{Esposito-Far{\`e}se}, G.}
\newblock \bibinfo{title}{{Tensor-scalar gravity and binary-pulsar
  experiments}}.
\newblock \emph{\bibinfo{journal}{\prd}} \textbf{\bibinfo{volume}{54}},
  \bibinfo{pages}{1474--1491} (\bibinfo{year}{1996}).
\newblock \eprint{gr-qc/9602056}.

\bibitem{hpk81}
\bibinfo{author}{{Haensel}, P.}, \bibinfo{author}{{Proszynski}, M.} \&
  \bibinfo{author}{{Kutschera}, M.}
\newblock \bibinfo{title}{{Uncertainty in the saturation density of nuclear
  matter and neutron star models}}.
\newblock \emph{\bibinfo{journal}{\aap}} \textbf{\bibinfo{volume}{102}},
  \bibinfo{pages}{299--302} (\bibinfo{year}{1981}).

\end{thebibliography}

%% Here is the endmatter stuff: Supplementary Info, etc.
%% Use \item's to separate, default label is "Acknowledgements"

\begin{addendum}
 \item The authors thank Paulo Freire for pointing out how useful PSR~J0337+1715 could be for testing the SEP.
The authors thank Lijing Shao for providing an independent cross-check on the signature of $\Delta$. The authors thank Kenneth Nordvedt for explaining why the signature of $\Delta$ differs from that in lunar laser ranging. A.M.A. is supported by an NWO Veni grant. N.V.G. is supported by NOVA and ASTRON. J.W.T.H.  acknowledges funding from an NWO Vidi fellowship and from the European Research Council under the European Union's Seventh Framework Programme (FP/2007-2013) / ERC Starting Grant agreement nr. 337062 (``DRAGNET''). A.T.D. is the recipient of an Australian Research Council Future Fellowship (FT150100415). S.M.R. is a CIFAR Senior Fellow. I.H.S. is supported by an NSERC Discovery Grant and by the Canadian Institute for Advanced Research. The NANOGrav project (involving D.L.K., D.R.L., R.S.L., S.M.R., and I.H.S.) receives support from National Science Foundation (NSF) Physics Frontiers Center award number 1430284. The National Radio Astronomy Observatory is a facility of the National Science Foundation operated under cooperative agreement by Associated Universities, Inc. The Green Bank Observatory is a facility of the National Science Foundation operated under cooperative agreement by Associated Universities, Inc.
\item[Software availability] All general-purpose software packages we use are open-source: psrchive (\url{http://psrchive.sourceforge.net/}),
TEMPO (\url{http://tempo.sourceforge.net/}),
TEMPO2 (\url{http://www.atnf.csiro.au/research/pulsar/tempo2/}),
numdifftools (\url{https://pypi.python.org/pypi/Numdifftools}),
sympy (\url{http://www.sympy.org/}), 
boost (\url{http://www.boost.org/}), galpy (\url{https://github.com/jobovy/galpy}), and GalDynPsr (\url{https://github.com/pathakdhruv/GalDynPsr}). The software written specifically for this project is also available at \url{https://github.com/aarchiba/triplesystem}
\item[Data availability] Data is available from the authors upon request.
 \item[Competing Interests] The authors declare that they have no
competing financial interests.
\item[Author contributions] A.M.A. wrote the processing pipeline, orbital modelling, fitting, and equation-of-state integration code; A.M.A. also ran the data processing and fitting operations. N.V.G. wrote the systematics analysis code. J.W.T.H. carried out an intensive observing campaign with the WSRT. A.M.A., N.V.G., J.W.T.H., D.R.L., R.S.L., S.M.R., and I.H.S. carried out observations with Arecibo and the GBT. J.W.T.H., S.M.R, and I.H.S. carried out a preliminary version of the data processing. A.T.D. consulted on the astrometry. D.L.K. and N.V.G. carried out the tidal-effects analysis. All authors participated in extensive discussions of the paper content.
 \item[Correspondence] Correspondence and requests for materials
should be addressed to A.M.A. (email: \href{mailto:a.archibald@uva.nl}{\nolinkurl{a.archibald@uva.nl}}).
\end{addendum}

%%
%% TABLES
%%
%% If there are any tables, put them here.
%%

\end{document}